\documentclass[12pt]{iopart}

\usepackage{iopams}
\usepackage{textcomp}
\usepackage{graphicx}
\usepackage{citesort}
\begin{document}

\title[Axion Helioscopes]{Axion Searches with Helioscopes and astrophysical signatures for axion(-like) particles}

\author{K~Zioutas$^{1,2}$, M~Tsagri$^1$\footnote{Present address: European Organization for Nuclear Research (CERN), CH-1211 Gen\`eve 23,
Switzerland}, Y~Semertzidis$^3$, T~Papaevangelou$^4$, T~Dafni$^5$
and V~Anastassopoulos$^1$}

\address{$^1$ University of Patras, Patras, Greece}
\address{ $^2$ European Organization for Nuclear Research (CERN), CH-1211 Gen\`eve 23,
Switzerland}
\address{$^3$ Brookhaven National Laboratory, NY-USA}
 \address{$^4$ IRFU, Centre d' \'{E}tudes Nucl´eaires de Saclay, Gif-sur-Yvette, France}
 \address{$^5$ Laboratorio de F\'isica Nuclear y Astropart\'iculas, Universidad de Zaragoza, Zaragoza, Spain}
\ead{\mailto{Thomas.Papaevangelou@cern.ch}}
\begin{abstract}
Axions should be produced copiously in Stars such as the Sun. The
first part of the article reviews the capabilities and performance
of axion helioscopes. The mechanism they rely on is described and
the achieved experimental results for the interaction of solar
axions and axion-like particles with matter are given. The second
part is actually observationally driven. New results obtained with
Monte Carlo simulation reconstruct solar observations, previously
dismissed, supporting an axion(-like) involvement with m$_{\rm
a}$$\approx$1-2$\times10^{-2}$\,eV/c$^{2}$. To further quantify
the suggested solar observations as being originated by axions,
additional theoretical work is needed. However, the recently
suggested axion interaction with magnetic field gradients is a
generic theoretical example that seems to reconcile for the first
time present limits, derived from axion helioscopes, and potential
axion-related solar X-ray activity, avoiding thus contradictions
with the best experimental limits. Magnetic quadrupoles can be
used to experimentally test this idea, thus becoming a new
catalyst in axion experiments. Finally, a short outlook for the
future is given, in view of the experimental expansion of axion
research with the state-of-the-art orbiting X-ray observatories.
\end{abstract}

\maketitle

\section{Motivation} \label{motivation}
The recent WMAP measurements \cite{Kom09} have established with
great precision that, about 84\% of the matter content of the
Universe is in the form of cold dark matter (CDM). The composition
of CDM is not yet known, however the most promising particle
constituents are WIMPs (Weakly Interacting Massive Particles) and
axions. In this work, we focus on the axions. The theoretically
introduced axions have far reaching consequences in astrophysics
and cosmology. At the theoretical level, the imprint of axions
appears in the QCD Lagrangian, which includes a CP-violating
parameter, the $\theta$-QCD ($\bar{\theta}$):

\begin{equation}
L_{CPV}= \bar{\theta} {\frac {\alpha_s} {8 \pi} G \tilde{G}} \,,
\end{equation}
\noindent where $a_s$ is the strong coupling constant and $G$ and
$\tilde{G}$ represent the gluon field and its dual. From this one
can estimate within an order of magnitude the neutron EDM
\cite{Bal79,Cre79}:

\begin{equation}
d_n(\bar{\theta}) \approx \bar{\theta} \frac {e} {m_n} \frac {m_*}
{\Lambda_{QCD}} \approx \bar{\theta} \cdot (5 \times 10^{-17}) \,
{\rm e \cdot cm} \,,
\end{equation}
\noindent with m$_{*}$= (m$_u$\,m$_d$)/(m$_u$+m$_d$)
being the reduced mass of the up and down quarks. $\Lambda_{\rm
QCD}$ is the QCD scale ($\sim$200\,MeV) and $m_n$ the neutron
mass. When the estimation is done more
precisely \cite{Khr00,Leb04,pospelov05,rob06} it comes out as
$d_n(\bar{\theta}) \approx \bar{\theta} \cdot (3.6 \times
10^{-16}) \, {\rm e \cdot cm}$.
The present neutron EDM limit \cite{neutron06} of $3 \times
10^{-26} \, {\rm e \cdot cm}$ results to a limit on $\theta$-QCD
of $ \bar{\theta} \leq 10^{-10}$. With the planned dEDM and pEDM
experiments at BNL, the experimental sensitivity can reach the
level of $\bar{\theta} \leq 10^{-13}$.

The value of $\bar{\theta}$ could potentially be between 0 and $2
\pi$. However, it turns out to be extremely tiny,
apparently due to some fine tuning mechanism. There is no symmetry
reason within the Standard Model for such a small value, creating
the so-called strong CP problem. Peccei and Quinn postulated that
such a tiny value arises from the breakdown of a new symmetry, which gives rise to axions
\cite{Pec77,Pec77a,Peccei:2006as,Weinberg77,Wil78}. The latter are
pseudoscalar particles arising as a solution to the CP problem in
strong interactions. They have properties closely related to those
of neutral pions. That is to say, the discovery of a QCD inspired
axion will explain why $\bar{\theta}$ is extremely small. Apart
from all this, axions also appear in string theory.

The standard axions were thought to have a symmetry-breaking scale
(or axion decay constant) of the order of the electroweak scale
\cite{Weinberg77,Wil78}. However, after such scenarios were ruled
out, new models were developed where that scale was arbitrary. In
fact, it could be so large that would make axions interact so
weakly that they were dubbed `invisible'. Such were the first,
thoroughly cited, axion models following the work of
Kim-Shifman-Vainshtein-Zakharov (KSVZ) \cite{Kim79,Shi80} and
Dine-Fischler-Srednicki-Zhitnitskii (DFSZ) \cite{Din81,Zhi80}.

In order to detect axions, one could rely on the generic property
of axions to couple to two photons \cite{Sikivie:1983ip}, as
described by the Lagrangian term

\begin{equation}
     {\cal L}_{{\rm a}\gamma\gamma}= -\frac{1}{4}\,g_{{\rm a}\gamma\gamma} F_{\mu\nu}\tilde F^{\mu\nu}a
     =g_{{\rm a}\gamma\gamma}\,{\bf E}\cdot{\bf B}\,a\,,
   \label{eq1}
\end{equation}

\noindent where $a$ is the axion field, $F$ is the
electromagnetic field-strength tensor, $\tilde F$ its dual, ${\bf
E}$ and ${\bf B}$ the electric and magnetic field, respectively.

Over the last three decades, different experimental techniques
have been developed for the search of the `invisible axions' or,
more generally, `axion-like particles': cavity searches for
galactic matter axions
\cite{Sikivie:1983ip,Bradley:2003kg,Hag98,Asz04,Duf05,Asz06},
solar axion searches using `helioscopes'
\cite{Sikivie:1983ip,Asz06,vanBibber:1988ge,Laz92,Mori98,Ino02,Ino08,Zio05,And07,Arik09},
the Bragg scattering technique
\cite{Paschos:1993yf,Avi98,Mor02,Ber01}, the polarization of light
propagating through a transverse magnetic field
\cite{Maiani:1986md,Raffelt:1987im,Cam93,Zav05}, photon
regeneration
\cite{Cam93,Ringwald:2003ns,Rabadan:2005dm,Kotz:2006bw,Lindner:2006,Cantatore:2006,Afanasev:2006cv,Afanasev:2006,Battesti:2006,Pugnat:2006,Sikivie:2007qm}
and others like the resonant method involving nuclear couplings
\cite{Mor95,Krc98,Nam07,Krc01,Lju04}\footnote{A simple comparison
between the methods, showing that the magnetic helioscopes are
more effective than the (crystalline) detectors, can be done by
calculating that (0.1\,T\,m)$^{2}$ corresponds to a detector
thickness of 2\,Z$^{-1}$\,kg/cm$^{2}$, where Z the atomic number
of the detector material.}.

The second part of the article is observationally driven. We
suggest atypical solar axion signatures as the so far unnoticed
manifestation of axions or axion-like particles (ALPs), which fit
decades-old puzzling solar observations and every day's experience
of solar X-ray telescopes. To better understand relevant
observations from the ubiquitously magnetized solar surface within
the axion or axion-like scenario, Monte Carlo simulations have
been performed, revising for the first time the propagation of
magnetically converted axions \cite{carlson} near the solar
surface, that is, the created X-rays. This is due to isotropic
Compton scattering of X-rays coming from axion conversion
underneath the lower chromosphere or even higher in the dynamic
atmosphere. For the conversion to happen efficiently, e.g. to
ensure a large axion-to-photon coherence length, it is reasonable
to assume that the axion rest mass must match the local plasma
density ($m_{\rm a}c^2 \approx \hbar \omega_{pl}$). To reconcile
magnetic field-related solar X-ray observations and the axion
scenario, the derived rest mass is, within a factor of 2-3,
$m_{\rm a}\approx 10$\,meV/c$^2$, which coincides with the upper
limit derived from SN1987A \cite{Raffelt:2006cw}, while
cosmological data allow for an axion rest mass below
$\sim$1\,eV/c$^{2}$ \cite{Han05}.

In fact, star evolution arguments exclude the involvement of
axions or the like (up to a certain coupling strength)
\cite{Raffelt:2006cw,Raf96}. In this work, we look more
specifically at solar X-ray observations as potential signatures
for new exotica, with preference to faint intensities, whose
luminosity cannot actually affect stellar evolution. With this,
however, we do not exclude a priori large flaring events as being
powered -or at least triggered- by solar exotica.

Other, recently published work
\cite{deA07,deA08,Ron09,Sim08,Fai09,Bur09}, refers to signatures
on very light ALPs (assuming m$_{\rm a}$$\ll$10$^{-7}$eV/c$^{2}$)
via the oscillatory behaviour of cosmic very high energy (VHE)
light and ALPs as they propagate over billions of years through
the intergalactic magnetic network. While the QCD-inspired axion
implies a particle with one rest mass and one coupling constant,
ALPs
 do not have to follow this constrain. As an
example, we mention the massive axions of the Kaluza-Klein type
\cite{DiL00}, which have a tower of mass states, but one
common coupling constant.

Throughout this work, references to the solar axion scenario
include any other particle candidate with similar properties
(ALPs).

\section{Solar axion flux}\label{axionflux}

\begin{figure}[htb!]
    \centerline{\includegraphics[width=0.8\textwidth]{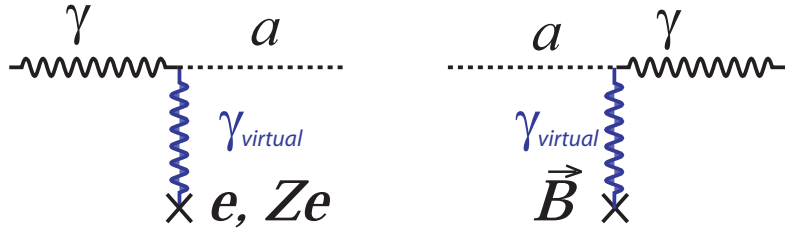}}
    \caption{Left: The incoherent Primakoff effect, which is assumed to occur inside the hot solar
    core and which gives rise to the creation of axions. Right: The inverse coherent process in
    a magnetic field, which is so far the working principle of an axion helioscope as it transforms the
    otherwise `invisible' axions to observable photons. This is an oscillation phenomenon,
    analogous to neutrino oscillations. The external magnetic field is needed in order to compensate the spin-mismatch in the case of photon-axion oscillation \cite{Ron09}.}
  \label{primakoff}
\end{figure}

\begin{figure}[t]
    \centerline{\includegraphics[width=0.7\textwidth]{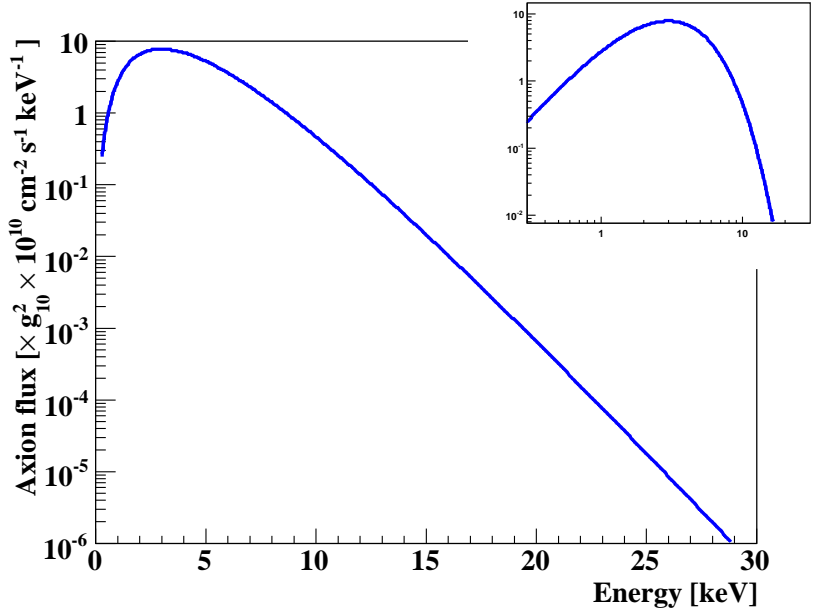}}
    \caption{The solar axion flux as expected at the Earth, following relation \ref{bestfit}, plotted in log-linear and log-log
 scales (inserted). The spectrum has a peak at around 3\,keV and mean energy of 4.2\,keV. The sub-keV range
 was experimentally not accessible before. Recently, first measurements have started covering the few-eV
 energy range \cite{giovanni}. Axion helioscopes are the best suited for low energy searches, since the
 screening effects, which appear in dense materials, are quasi-suppressed.}
  \label{axion-flux}
\end{figure}

The Lagrangian term (\ref{eq1}) points to the ${\bf E}\cdot{\bf B}$ in a
hot thermal plasma as a source for the
production of axions. {\bf E} is provided by the charged particles of
the plasma whereas {\bf B} arises from propagating thermal photons. The
Primakoff process describes an incoherent production, where a
photon converts into an axion in the electric field of a charged
particle (Figure \ref{primakoff}).
Figure \ref{axion-flux} shows the axion flux spectrum as expected
at the Earth; it is, essentially, a blackbody distribution of the
thermal conditions in the solar interior, which presents a maximum
at the energy of 3\,keV and a mean energy value of $\langle E
\rangle$=4.2\,keV. The reason for the higher energy than the one
in the Sun hot core (kT$\sim$1\,keV) is the suppression of low
energies (large wavelengths) which reaches
in total a factor of $\sim$25 due to screening effects \cite{Raffelt:1985nk}.
An analytic approximation of the flux is given by \cite{And07}:

\begin{equation}\label{bestfit}
  \frac{{\rm d}\Phi_{\rm a}}{{\rm d}E}=6\times
  10^{10}~{\rm cm}^{-2}~{\rm s}^{-1}~{\rm keV}^{-1}\, g_{10}^2
  \,E^{2.481}{\rm e}^{-E/1.205}\,
\end{equation}

\noindent where $\Phi_{\rm a}$ is the axion flux and
$g_{10}=g_{{\rm a}\gamma\gamma}/10^{-10}~{\rm GeV}^{-1}$. This
equation takes into account the Primakoff effect of real thermal
photons interacting with the Coulomb electric field of the solar
core plasma in atomic scale. Most of the axion flux emerges from
$R\lesssim0.1R_{\odot}$ or from $\sim$2\% of the solar disk
surface (for an earth-bound observer).

So far, the impact of the solar magnetic fields has been widely
ignored. This point will be addressed in (sections \ref{sun} and
\ref{xray}). Some fine tuning cannot be excluded, taking into
account the variations of the dynamic Sun in all spatio-temporal
scales, in particular in the outer layer
(R$\gtrsim$0.9\,R$_\odot$). The inner solar magnetic field must
reach approximately 50 to 100\,T at the bottom of the convection
zone (about 200\,000\,km below the surface, i.e. at
R$\simeq0.7\,R_\odot$), while even much higher fields might exist
deeper in the Sun \cite{Cou03}.

\section{Principles of Detection}\label{detection}

The search for solar axions with helioscopes is based on the
inverse coherent Primakoff effect (Figure \ref{primakoff}): solar axions
coming from the Sun (produced via the incoherent Primakoff effect)
will be re-converted to X-ray photons as they pass through a
(strong) transverse laboratory magnetic field. These excess
photons would then be seen in X-ray detectors
\cite{Sikivie:1983ip}, located outside the actual magnetic field
region. The number of photons expected to reach these detectors is
\begin{equation}\label{nphotons}
N_{\gamma} = \int \frac{{\rm d}\Phi_{\rm a}}{{\rm d}E}\,P_{{\rm a}\to\gamma}\,A\,{\it t}\,{\rm d}E
\end{equation}
\noindent where ${\rm d}\Phi_{\rm a}/{\rm d}E$ is the axion
spectrum as expected at the Earth (\ref{bestfit}), $P_{{\rm
a}\to\gamma}$ the probability of the axion-to-photon conversion,
${\it t}$ the observation time and $A$ the axion-sensitive area of
the magnet aperture. The conversion probability is expressed as
\begin{equation}\label{convProb}
P_{{\rm a}\to\gamma}=\left(\frac{g_{{\rm a}\gamma}\,B}{2}\right)^{2}\,\frac{1}{q^2+\Gamma^{2}/4}\,[1+e^{-\Gamma L}-2e^{-\Gamma L/2}\,cos{qL}]\,\,.
\end{equation}
With $\Gamma$, the inverse
absorption length in the medium, the general case of the presence of a refractive medium
inside the magnetic pipes has been included. The axion-photon momentum transfer $q$ is given by $q=(m_{\rm
a}^2-m_{\gamma}^2)/(2E_{\rm a})$ and m$_{\gamma}$ is the effective mass of the photon as
acquired due to the medium (see (\ref{mgamma})). Substituting
(\ref{convProb}) and (\ref{bestfit}) in (\ref{nphotons}), we get

\begin{equation}\label{eq5}
\fl N_{\gamma} \simeq 10^{-6}~{\rm cm^{-2}~s^{-1}~keV^{-1}}\,
 g_{10}^4\, E^{\,2.481}{\rm e}^{-E/1.205}
 \left(\frac{L}{10~\rm m}\right)^2
 \left(\frac{B}{9.0~\rm T}\right)^2\,A\,{\it t}\,{\rm d}E \nonumber \,.
\end{equation}
\noindent This equation makes evident the importance of the factors
$B$, $L$ and $A$ in the detection of axion-converted photons.
 To maintain the maximum conversion probability, i.e. zero momentum transfer ($q \rightarrow 0$), the axion and photon fields
 need to remain in phase over the length of the magnetic field. This coherence condition is met when
 $qL \lesssim \pi$, meaning that the experiment is sensitive to different mass ranges depending on q:
\begin{equation}\label{mageneral}
\sqrt{m_{\gamma}^{2} - \frac{2\pi E_{{\rm a}}}{L}} \lesssim m_{{\rm a}} \lesssim \sqrt{m_{\gamma}^{2} + \frac{2\pi E_{{\rm a}}}{L}}
\end{equation}

\noindent The fractional resolution can be written as:
\begin{equation}\label{width}
\frac{dm_{\rm a}}{m_{\rm a}}\equiv
\frac{dm_{\gamma}}{m_{\gamma}}=\frac{2\pi\,E_{\rm a}}{L\,m_{\rm
a}^2}.
\end{equation}

\noindent Taking an axion energy of $E_{\rm a}$=1\,eV,
m$_{\gamma}$=10$^{-3}$\,eV/c$^{2}$ and $L$=10\,km, the width
becomes
\begin{equation}
\frac{{\rm FWHM}}{m} \simeq 1.2\times10^{-4}.
\end{equation}

\noindent Further, for solar axions with E$_{\rm a}$=4.2\,keV,
$m$=1\,eV/c$^{2}$ and $L$=4\,m \cite{vanBibber:1988ge},
\begin{equation}
\frac{FWHM}{m} \simeq
10^{-3}\left(\frac{m_{\gamma}}{eV}\right)^{-2} \,,
\end{equation}
\noindent for $m_{\gamma}$=0.1\,eV/c$^{2}$, FWHM=10\%.

\noindent When changing to fractional density, one obtains
\begin{equation}\label{rho}
\frac{d\rho}{\rho}=2\frac{dm}{m} \,.
\end{equation}

\noindent Choosing E$_{\rm a}$=4.2\,keV and $m_{\rm
a}$=10$^{-2}$\,eV/c$^{2}$, the allowed density fluctuations should
not be more than d$\rho / \rho \simeq 10^{-2}$, in order to
maintain the coherence effect over the whole length. For
the static Sun these conditions are encountered at a few hundreds
of km underneath the surface and the coherence length there can be
$L \simeq$ 3\,km (to be compared, for example with the CAST 10\,m
length)\footnote{Taking into account the Sun's dynamical character
near its surface it is still reasonable to assume, as a numerical
example, $L$$\approx$20\,km and $B$$\approx$1\,T \cite{Liv06}. This
results to $P_{{\rm a}\to\gamma}$$\approx$10$^{-12}$
($g_{a\gamma\gamma}=g_{10}$). For comparison, the surface X-ray
brightness of a large flare requires instead a conversion
efficiency of $\gtrsim10^{-3}$. Then, the brightness of such
events cannot be explained quantitatively only by QCD-axions using
the inverse Primakoff effect, even assuming favourable input
parameters. We do not reject these events from further
consideration, since the rest of their behaviour fits the axion
scenario of this work. Nevertheless, much lower X-ray
brightenings, like microflares, nanoflares, etc., but also the
quiet Sun itself, should then be more appropriate for the
reasoning of this work (see \sref{yohkoh}).}.

\subsection{Data-taking strategies}\label{datataking}
\begin{figure}[htb!]
    \centerline{\includegraphics[width=0.5\textwidth]{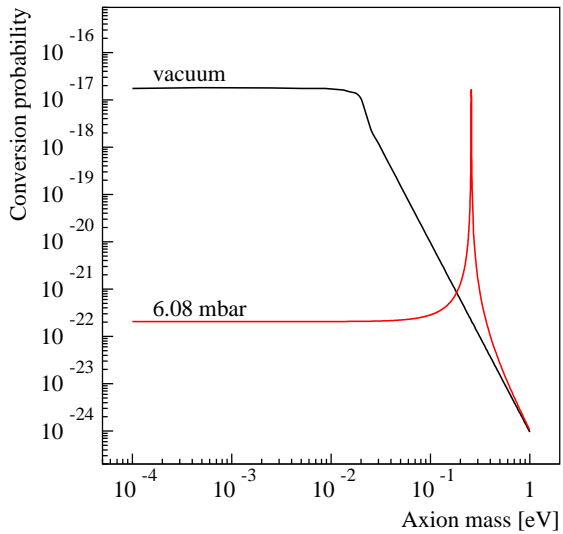}}
    \caption{The axion-to-photon conversion probability versus the axion rest mass,
    assuming an axion-photon coupling constant of $g_{a\gamma\gamma}$=1$\times10^{-10}$\,GeV$^{-1}$.
    In black, the line corresponding to vacuum inside a 10\,m long magnetic pipe at 9\,T.
    For $m_{\rm a}$$\geq$0.02\,eV/c$^{2}$, the conversion efficiency breaks down because the incident axion
    and the emerging photon waves get out of phase (deconstructive interference). In red, the conversion
    probability for a specific Helium density setting (equivalent to 6.08\,mbar at 1.8\,K).
    The shown resonance curve has a very narrow ($\sim$0.5\%) width for which the specific pressure (density)
    of the refractive material (Helium gas)
    restores the coherence over the whole length \cite{And07}.}
  \label{vac_gas}
\end{figure}

\noindent {\bf Vacuum} In the particular case in which the magnetic field
where axions are converted into photons is under vacuum
($\Gamma \to 0, m_{\gamma}\to 0$), equation (\ref{convProb})
becomes \cite{vanBibber:1988ge}
   \begin{equation}
      P_{{\rm a}\to\gamma}=\left( \frac{g_{{\rm a}\gamma\gamma}B}
          {q} \right)^2 \sin^2\left( \frac{qL}{2} \right)\,,
    \label{vacuum1}
   \end{equation}
    while $q=m_{\rm a}^{2}$/2E$_{\rm a}$. Applying the coherence condition, the range of axion rest masses one is sensitive to is
    \begin{equation}\label{vacuum2}
        m_{\rm a} \lesssim \sqrt{\frac{2\pi E_{\rm a}}{L}} \,.
    \end{equation}
\noindent This implies that for a magnetic field of $L$=10\,m and
for a mean solar axion energy of 4.2\,keV, the sensitivity of an
experiment would cover axion rest masses of $m_{\rm a}
\lesssim$0.02\,eV/c$^{2}$
(Figure \ref{vac_gas}).\\
{\bf Refractive gas} More massive axions will begin to fall out of
phase with the emerged photon waves, due to different velocities.
The addition of a refractive material (a buffer gas, for example)
will `slow' the photon giving it an effective mass (m$_{\gamma}$).
In order to compensate for the velocity mismatch and restore
maximum conversion probability, m$_{\gamma}$ should be adjusted to
approach m$_{{\rm a}}$. In a low-$Z$ gas, where self absorption is
minimum, the effective photon mass (m$_{\gamma}$) and the axion
rest mass must fulfill the following condition
\cite{vanBibber:1988ge}
    \begin{equation}\label{mgamma}
      m_{\rm a}\,c^{2} = m_{\gamma}\,c^2= \hbar \,\omega_{pl} = \hbar\,\sqrt{\frac{4 \pi \alpha n_{\rm e}}{m_{\rm e}}}\,
      ,
    \end{equation}
\noindent in which $\alpha$ is the fine structure constant and
$m_{e}$ and $n_{e}$ the electron mass and number density
respectively \footnote{Although in the actual experiments the
refractive gases are not ionized, this equation holds as long as
the axion/photon energies involved are much higher than the
binding energy of the atomic electrons of the refractive gas.
However, one needs to exercise cautiousness when applying this
method searching for much lower energy solar axions, if their
energy approaches that of atomic transitions.}. For a rest mass of
1\,eV/c$^{2}$, this would imply a required pressure at room
temperature of almost 15\,bar or, for the case of Helium in the
cryogenic environment of a magnet (e.g. 1.8\,K), a pressure of
90\,mbar. Extending the calculations to the Sun surface, one finds
$\hbar\,\omega_{pl}\simeq 10^{-2}$\,eV for
$\rho=2\times10^{-7}$\,g/cm$^{3}$, or $\hbar\,\omega_{pl}\simeq$
300\,eV in the core ($\rho$=150\,g/cm$^{3}$).

Different axion masses can be tuned by changing the gas density
(pressure) in discrete steps. For each density, the coherence
condition is restored, but only for a very narrow mass range
around $m_{\gamma} = m_{\rm a}$. For m$_{\rm a}$=1\,eV/c$^{2}$,
$L$=10\,m and a mean axion energy of E$_{{\rm a}}$= 4.2\,keV,
(\ref{width}) and (\ref{rho}) imply a required constant density at
a level of d$\rho$/$\rho\simeq$10$^{-3}$. An example of the
probability conversion for two different pressures (vacuum and
6.08\,mbar) is given in Figure \ref{vac_gas}. To assure density
homogeneity over the length of the magnetic pipes, even gravity
effects have to be taken into account when tracking the sun with
the magnet tilted. The effect of the gravity in the gas density in
the magnetic cold bores along with the attenuation of the X-rays
are the ultimate limits for the performance of an axion helioscope
(for a detailed discussion, see \cite{Cre08}).

Another strategy of taking data with gas in the pipes is by performing a
continuous `scanning' over a large range of pressures during one solar tracking
measurement. The advantage this approach provides with respect to
the previous one is that it allows for a `fast-look' to a wide
range of axion rest masses, but with lower sensitivity. A possible
signal candidate can be scanned later over longer time intervals,
following a specific, predefined protocol procedure, excluding in
this way any kind of bias in the selection. A similar `scanning',
but covering eventually a much wider density [=axion rest mass] range,
is suggestive of taking place in the dynamic Sun (see \sref{sun}).

\section{Axion Identification Techniques}\label{idtechniques}

In this section we discuss three techniques that can be
applied for the identification of an unambiguous solar axion signal, either
individually or combined:

\noindent {\bf Excess} Since axions are expected to oscillate into photons only
while traversing a dipole magnetic field which is pointing at the
centre of the Sun, one expects an excess in X-rays compared to the
periods when the Sun is out of the field of view of the magnet
(background). In the case of a signal candidate one has -in
principle- the possibility to change the magnetic field, repeat
the run and investigate whether the candidate signal follows the
$B^2$ dependence dictated by (\ref{eq5}).
This is usually considered as the ultimate cross check, although
increasing the field is hardly an option (given that one usually
already runs at the maximum magnetic field, and running the magnet
with significantly reduced field is not trivial either).

\begin{figure*}
  \begin{minipage}{0.49\textwidth}
      \centerline{\includegraphics[width=0.99\textwidth]
                                  {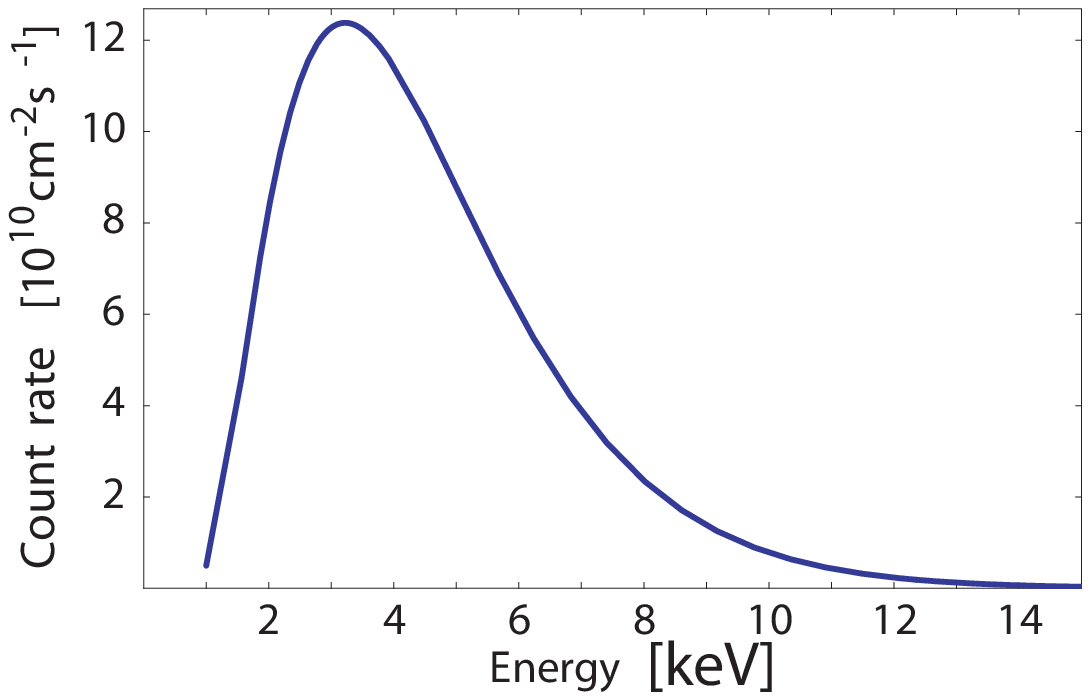}}
  \end{minipage}
  \begin{minipage}{0.49\textwidth}
      \centerline{\includegraphics[width=0.99\textwidth]
                                  {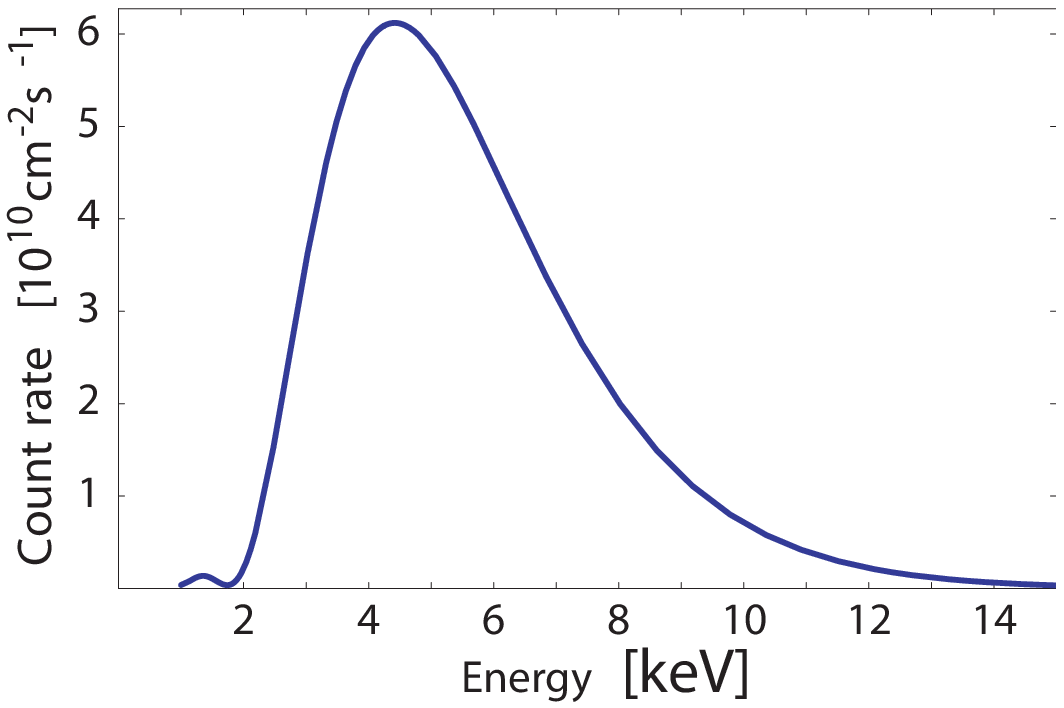}}
  \end{minipage}
  \begin{minipage}{0.49\textwidth}
      \centerline{\includegraphics[width=0.99\textwidth]
                                  {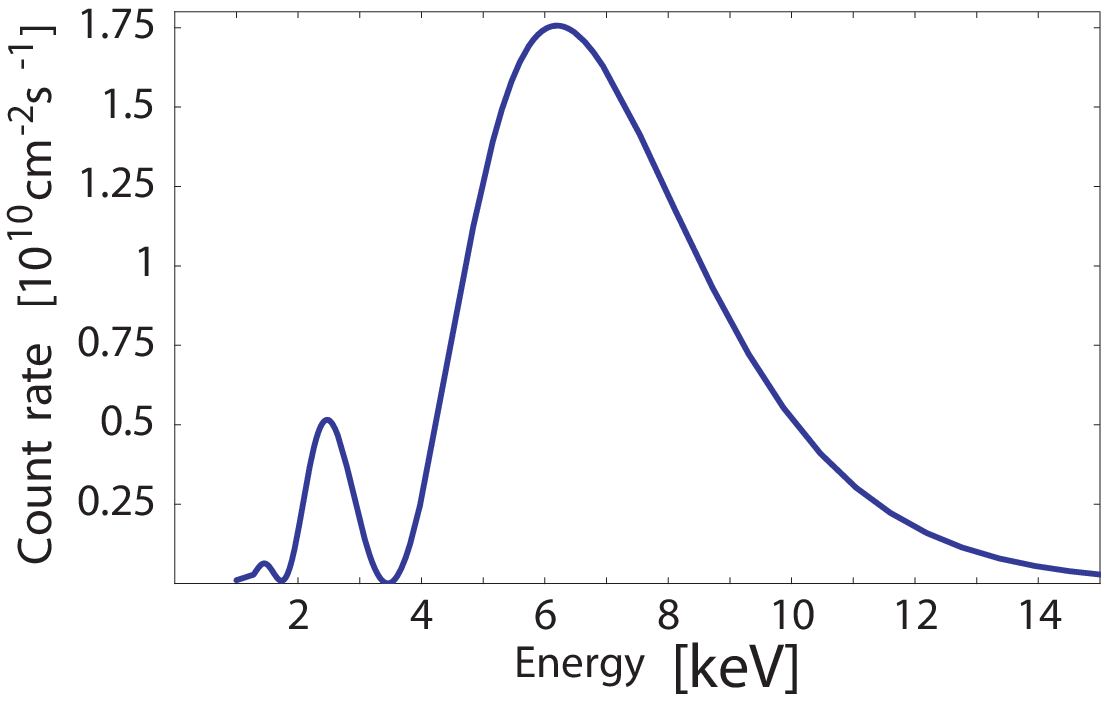}}
  \end{minipage}
  \begin{minipage}{0.49\textwidth}
      \centerline{\includegraphics[width=0.99\textwidth]
                                  {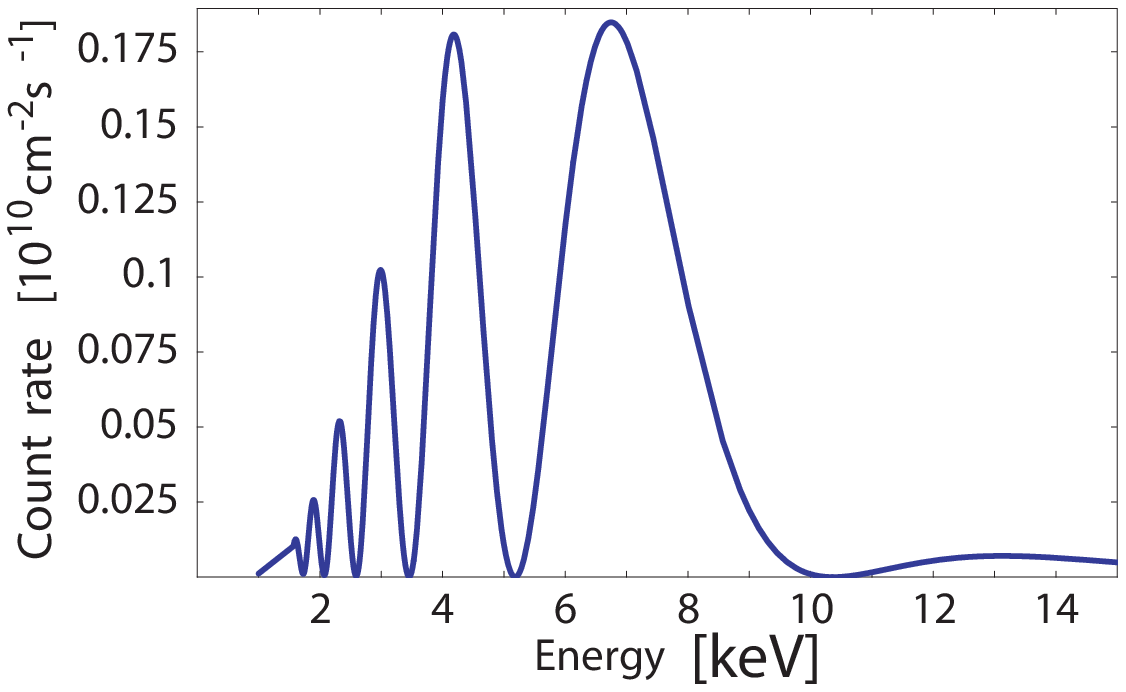}}
  \end{minipage}
  \caption{Expected analog photon spectra (assuming an axion-photon coupling constant
  $g_{\rm a \gamma\gamma}$=1$\times10^{-10}$\,GeV$^{-1}$) depending on the shift
  S=m$_{\gamma}$-m$_{a}$ from the resonance: S=0 (top left), S=0.5$\times$FWHM (top right),
  S=FWHM (bottom left) S=3$\times$FWHM (bottom right) \cite{Arik09}.
  Note the change of the spectral shape and decrease in intensity with increasing S.} \label{offresonance}
\end{figure*}

\noindent {\bf On/Off-resonance identification technique} This
technique allows to definitely and precisely establish a potential
axion signal and its rest mass, provided that m~$_{\rm a}$$\gtrsim$~
0.01\,eV/\,c$^{2}$, for $L$=10\,m. In addition, the CAST
collaboration \cite{Arik09} for the first time explains how the
spectral distribution of the axion-converted photons depends on
the momentum mismatch between the axion and the emerging
``massive'' photon m$_{\gamma}$ (Figure \ref{offresonance}). The
striking oscillatory behaviour of the otherwise smooth solar axion
spectrum provides an undoubtable signature.

\begin{figure*}[t]
    \centerline{\includegraphics[width=0.7\textwidth]{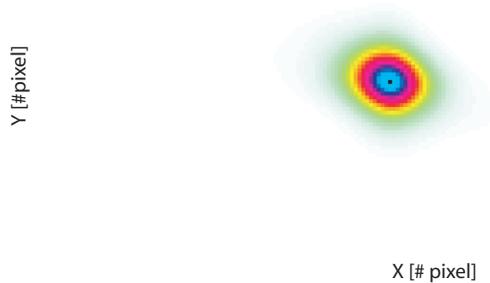}}
    \caption{Simulation of the expected axion image of the Sun focusing the coherently converted
    axions-to-X-rays inside the magnetic field of CAST on the CCD pixel
    detector (64$\times$200 pixels of 150\,$\mu$m$\times$150\,$\mu$m)\cite{And07}.}
\label{telescope}
\end{figure*}
\noindent {\bf Focusing devices} The discovery potential of an
axion helioscope is improved significantly by using an X-ray
focusing device, as it has been implemented for the first time by the CAST collaboration.
 An X-ray telescope (or other focusing optics) presents a threefold
importance: a) it projects all the axion-converted X-rays entering
from the axion sensitive region (in general, the magnet aperture
with a surface of tens of cm$^{2}$) onto a small spot of few
mm$^{2}$ at the focal plane, thus increasing the S/B ratio by 1 or
2 orders of magnitude, b) it allows for a \textit{simultaneous}
measurement of signal (inside the spot area) and background
(outside the expected spot), a unique possibility in general and
c) as an imaging device, a strong signal resulting from
axion-to-photon conversion will reflect the energy and radial
intensity distribution of the inner Sun (Figure \ref{telescope}),
with an unprecedented accuracy, having in mind for comparison the
poor reconstruction with the solar neutrinos.

\noindent We point out that the working principles of both
techniques, i.e. the excess as well as the on/off-resonance ID, apply
to the Sun's huge-sized magnetic fields. In fact, we use
below both to understand otherwise unexplained and unpredictable
spatio-temporally varying solar X-ray surface brightness.
\section{Axion Helioscopes}\label{helioscopes}

In the following we describe the short list of the axion
helioscopes built up to now.

\begin{figure*}
  \begin{minipage}{0.49\textwidth}
      \centerline{\includegraphics[height=0.65\textwidth]
                                  {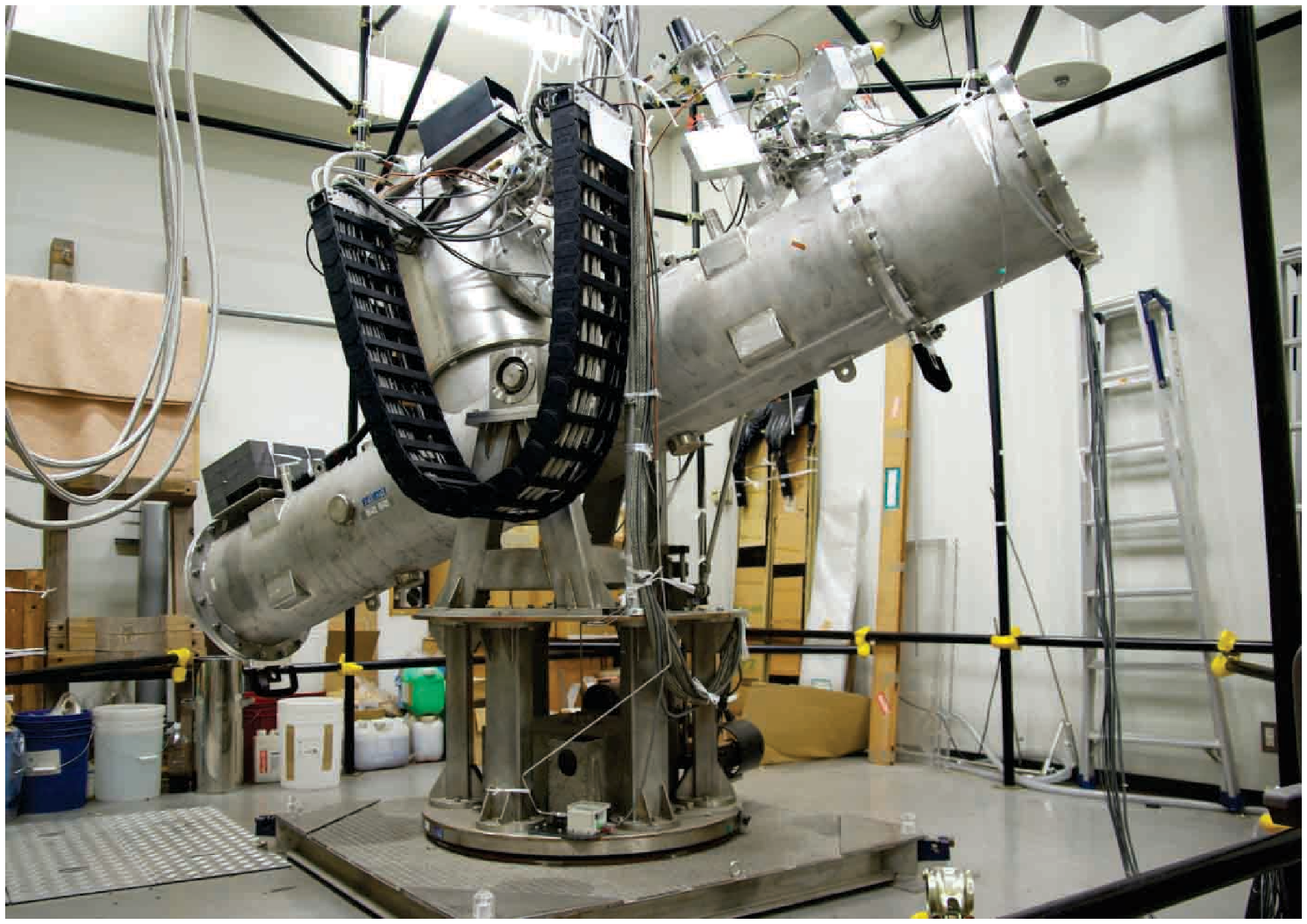}}
  \end{minipage}
  \begin{minipage}{0.49\textwidth}
      \centerline{\includegraphics[width=0.99\textwidth]
                                  {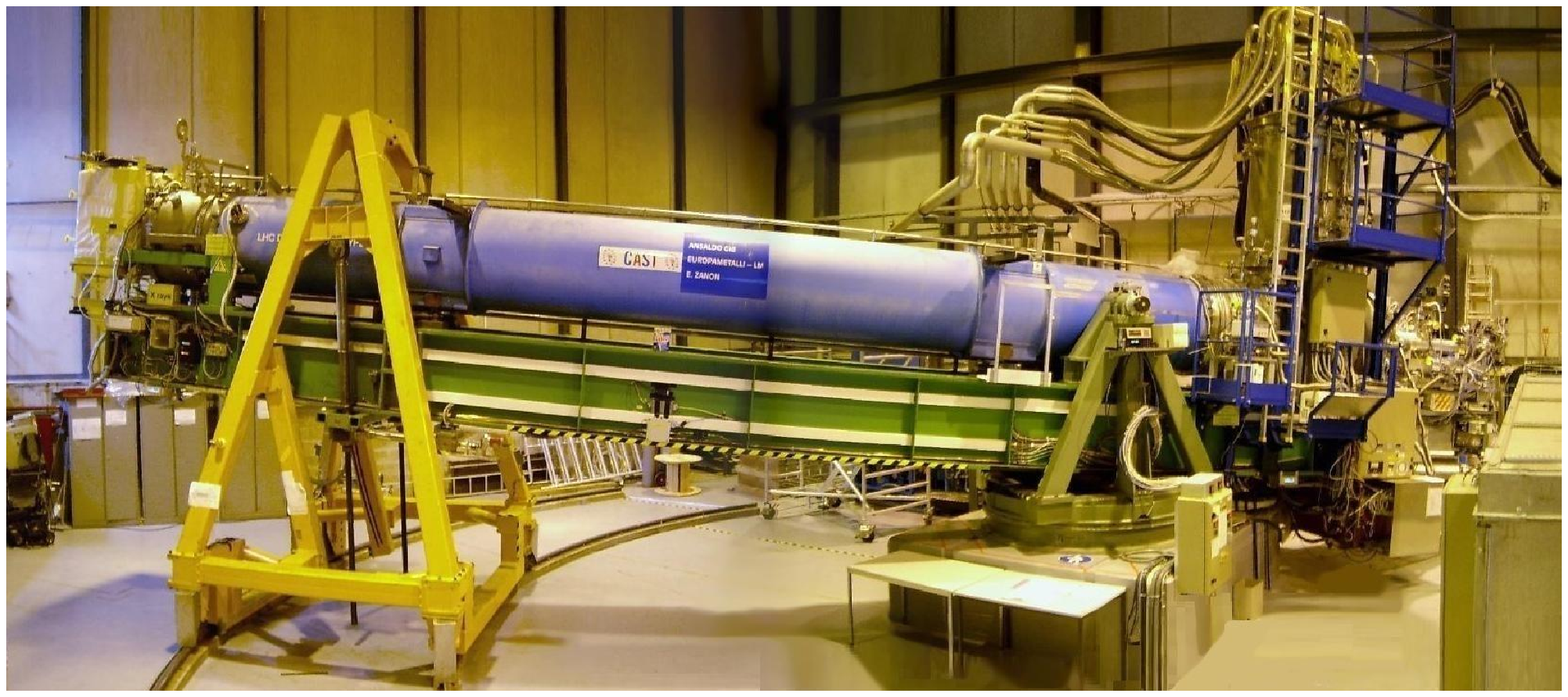}}
  \end{minipage}
  \caption{Recent pictures of the active axion helioscopes, Sumico on the left and CAST on the right.}
\label{pictures}
\end{figure*}
\begin{figure*}[htb!]
\begin{center}
 \centerline{\includegraphics[width=0.65\textwidth]{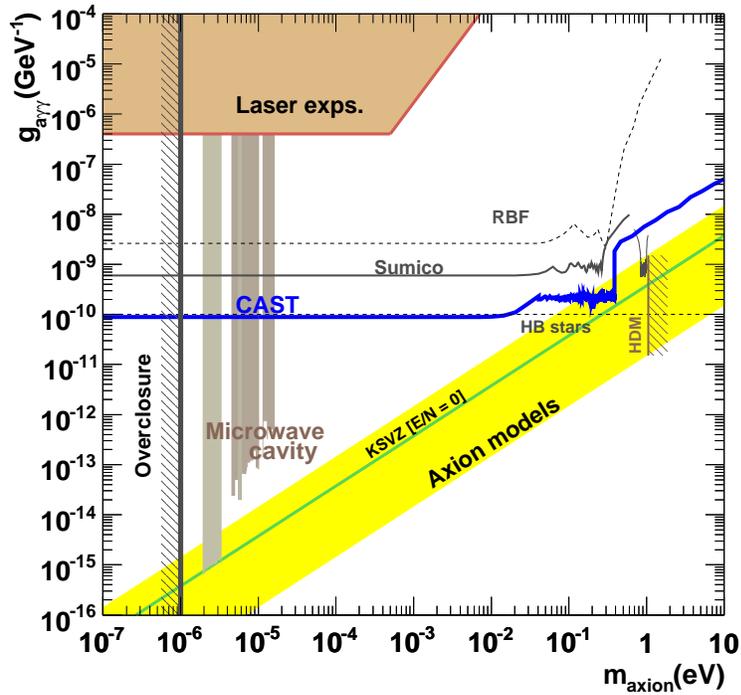}}
 \caption{
   Exclusion plots in the axion-photon coupling versus the rest mass of the QCD-inspired
   axions. The limits achieved by helioscopes (RBF~\cite{Laz92},
   Sumico~\cite{Mori98,Ino02,Ino08}, CAST \cite{Zio05,And07,Arik09})
   are put in the general picture of the axion searches. For comparison are given the
   astrophysically and cosmologically derived conclusions: HB
   stars~\cite{Raffelt:2006cw,Raf96}, the
   hot dark matter limit (HDM) for hadronic axions
   $m_{\rm a}$$<$1.05\,eV/c$^{2}$~\cite{Han05} inferred
   from WMAP observations of the cosmological large-scale structure,
   and the lower rest mass limit following overproduction of dark matter axions
   (overclosure of the Universe).
   \label{exclusion}}
 \end{center}
\end{figure*}

\subsection{The pioneering axion telescope}\label{rbf}
The first axion helioscope was built in 1992 by the
Rochester-Brookhaven-Fermi collaboration \cite{Laz92}, following
the recipe given in \cite{Sikivie:1983ip}. Their `axion converter'
was a 2.2\,T dipole magnetic field extended over 1.8\,m. This magnet,
lying fixed, was oriented towards the setting of the Sun in such a
way that it would actually point directly to the Sun during
$\sim$15\,min for a few days. For the detection of the converted axions
an X-ray proportional chamber was employed. Data with this
pioneering solar axion device were taken for about 4\,h with the
magnet pipe in vacuum and another 4\,h with Helium buffer gas in
the pipe in two different pressures. The results (Figure
\ref{exclusion}) were more sensitive than the laser experiments of that
time by roughly two orders of magnitude, but still far
from the theoretically motivated line (QCD axions).

\subsection{The Tokyo helioscope (Sumico)}\label{tokyo}
In 1997, in the University of Tokyo, a second generation axion
helioscope took data, Sumico (Figure \ref{pictures})\cite{Mori98}.
A 2.3\,m long magnet which reaches a field of 4\,T, was mounted on
a platform that enabled it to follow the Sun for approximately
half a day (!). PIN photodiodes were used as X-ray detectors. Data were
taken for one week with the magnet bores under vacuum conditions. The system was upgraded in 2000, when for one month
measurements were taken with Helium gas in the magnetic pipes.
When combining these two periods of measurement, an upper limit on
the axion-to-photon coupling constant was derived for $m_{\rm
a}$$<$0.27\,eV/c$^{2}$ \cite{Ino02}. Recently, the Sumico collaboration
presented results for 34 mass settings around 1\,eV/c$^{2}$.
These measurements give the most stringent limit, so far, on the
axion coupling for masses $0.84<m_{\rm a}<1.00\,$eV/c$^{2}$
(Figure \ref{exclusion}) \cite{Ino08}.

\subsection{The CERN Axion Solar Telescope (CAST)}\label{cast}
CAST \cite{Zio05,And07,Arik09,giovanni} represents the third
generation of axion helioscopes (Figure \ref{pictures}). In fact,
it outscores the previous ones in many of the important
parameters, starting with the key feature, the magnet, as can be
seen in \Tref{figmerit}.
It uses a decommissioned LHC prototype magnet which reaches a field of 9\,T
inside two parallel pipes of length 9.26\,m and aperture
14.5\,cm$^{2}$ each. This feature makes CAST the most sensitive
helioscope built so far. The magnet is mounted on a moving
platform which allows it to follow the Sun for approximately
2$\times$1.5\,h per day.
Currently CAST uses 3 Micromegas X-ray detectors and an X-ray
mirror optics coupled with a CCD camera at its focal plane (X-ray
telescope). The latter system distinguishes CAST's performance from any other
axion helioscope (see \sref{idtechniques}). The
coherence condition restricts the CAST sensitivity to $m_{\rm
a}\leq$ 0.02\,eV/c$^{2}$ when the magnetic pipes are under vacuum
(Figure \ref{vac_gas}). The CAST result is, up to now, the most
restrictive experimental limit on the axion-photon coupling constant for this mass
range. Moreover, it competes with the previous astrophysical
limit based on the Helium-burning lifetime of HB stars.
Subsequently, CAST extended its sensitivity to m$_{\rm a} \leq
$0.4\,eV/c$^{2}$ using $^4$He inside the magnet pipes. Replacing the $^4$He by $^3$He, CAST
can cover the axion rest mass range up to $\sim$1.2\,eV/c$^{2}$.\\

\noindent According to (\ref{eq5}), the sensitivity of an axion
helioscope is determined by the following parameters: the strength
of the magnetic field $B$, its length $L$, the effective
axion-sensitive magnetic aperture $A$ and the time of measurement
$t$. \Tref{figmerit} shows a comparison of the axion helioscopes
based on these characteristics. The figure of merit, the
product $(BL)^{2}$, which plays the most important role, is
given in the first column. The next columns enhance the
comparison with the additional information of the aperture and the
axion exposure time. Sumico and CAST are the first `direct-search'
experiments that can probe the QCD axion model strip near the eV
range in the $g_{\rm a\gamma\gamma}$-m$_{\rm a}$ parameter space.
\begin{table}
\caption{\label{figmerit} Comparison of figures of merit of the
axion helioscopes. $B$ is the strength of the magnetic field, $L$
its length, $A$ the effective, axion-sensitive magnetic aperture
and $t$ the tracking time per day (for the orbiting telescopes see
\cite{carlson,Dav06}).}
\begin{indented}
\lineup \item[]\begin{tabular}{@{}llll} \br
Helioscope& $(BL)^2$ & \0\0\0$(BL)^2$\,$A$ & $(BL)^2$\,$A$\,$t$\\
 & T$^2$\,m$^2$ & \0\0\0T$^2$\,m$^4$ & T$^2$\,m$^4$\,hours\\
 \mr
RBF &\0\016&$\0\sim$\03$\times10^{-2}$&$\0\sim$\0\01$\times10^{-2}$\\
Sumico &\0\085&$\0\sim$10$\times10^{-2}$&$\0\sim$120$\times10^{-2}$ \\
CAST & 6946&\02000$\times10^{-2}$&\0\06000$\times10^{-2}$\\
In orbit &\0324 &20000$\times10^{-2}$ &\0\0\0\0\0--\\
\br
\end{tabular}
\end{indented}
\end{table}

\subsection{Orbiting X-ray telescopes }\label{orbit}
The same working principle as the one of the above mentioned axion
helioscopes can be applied with orbiting detectors sensitive to
hard X-rays (see for example \cite{Han07,rhessi2} for the case of
the RHESSI solar X-ray observatory, with a threshold above
$\sim$3\,keV). Axion-to-photon conversion may occur either in the
terrestrial magnetic field \cite{Dav06}, or in the one near the
solar atmosphere \cite{carlson}. Despite the big differences
between the two schemes, their sensitivity can compete with the
best earth bound helioscope (see next section), but only for an
axion rest mass range (far) below $10^{-4}$eV/c$^2$. Furthermore,
following the present article, solar X-ray telescopes in space may
operate as the most sensitive solar axion antennas for more
massive axions (m$_{\rm a}>10^{-3}$eV/c$^2$) by utilizing,
complementarily, the solar magnetic fields near the photosphere.
One is inclined to (Erroneously) consider such a scheme as an
indirect method in axion helioscopy. The so obtained results can
be equally significant, and eventually with a better built-in
sensitivity, compared to the direct axion-detection techniques
using axion helioscopes.

\section{Photosphere: the resonant-coherent axion-photon converter?} \label{sun}

\begin{figure*}[htb!]
      \centerline{\includegraphics[height=0.49\textwidth]
                                  {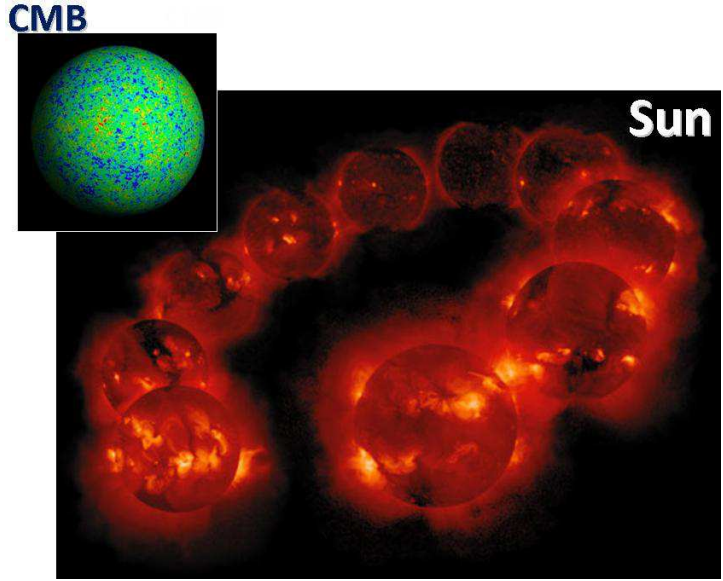}}
  \caption{The changing brightness of the Sun in soft X-rays in different
  periods of its mysterious 11-year cycle was measured with the X-Ray
  Telescope of the Japanese Yohkoh mission (1991-2001). The solar atmospheric
  temperature fluctuations are huge ($\Delta T/T$$\sim 10^{3}$). The (quiet) Sun
  X-rays are conservatively unexpected, and this is the solar coronal heating problem (since 1939),
  which remains {\it ``one of the most perplexing and unsolved problems in astrophysics to date''} \cite{Ana09}.
For comparison, in the insert (upper left) the temperature
fluctuations $\Delta T/T$ of the infant cosmic plasma are shown
(at the $10^{-5}$ level, as they have been measured by WMAP). The
temperature of both is similar: 5800\,K versus 3000\,K. One
difference to be noticed is the quasi zero magnetic field in the
cosmic plasma versus the varying solar magnetic fields in the
Tesla scale. Remarkably, solar X-ray emission, above its steady
component, follows spatio-temporally magnetic activity. Courtesy
M.~DiMarco/HAO \& NCAR \cite{irradref}. }
   \label{suns}
\end{figure*}

In this second part, we move from the short man-made to large,
natural occurring axion-photon converters near the solar surface.
A rather large coherence length ($>$1-10\,km) can be at work
thanks to the low density and low Z solar gas. Indeed, in a specific
layer of the continuously varying density, the
resonance condition $\hbar \omega_{pl} \approx m_{\rm a}c^{2}$ can restore coherence,
provided the axion rest mass is above $\sim$meV/c$^2$.
 A first, rough comparison between the temperature of the solar plasma
($\sim$ 5800\,K) and the well studied one of the infant Universe
(Figure \ref{suns}, upper left), of a rather similar temperature
($\sim$3000\,K), is interesting due to the striking contrast
between the perfect cosmic blackbody distribution and the
equivalent one from the Sun (Figure \ref{black}). To put it
differently: if the predicted and measured tiniest fluctuations of
the cosmic plasma of $\Delta T/T \sim10^{-5}$ provide(d)
fundamental new physics, one is even more tempted to conclude that
the unpredictable and huge solar atmospheric fluctuations ($\Delta
T/T\sim10^3$) might be the imprints of hidden new physics.

There is a fundamental difference between both plasmas under
comparison: it is only the solar plasma, which is permeated with
unpredictable, huge-sized and $\sim$Tesla strong magnetic fields.
Is this already a hint for axions, following the previous
sections? We follow this simplified question in this second part
of the article, which is observationally motivated.

\subsection{General considerations}\label{considerations}
In the following, firstly, we suggest atypical solar axion
signatures, even though the observed strong X-ray intensities
cannot be reproduced rigorously within the QCD-inspired axion
picture, i.e. only via the coherent inverse Primakoff effect; this
is suggestive of other exotica with similar properties, or another
axion-to-photon conversion process (our preference at present), or
both. The Primakoff effect is at present the process behind the
working principle of almost all axion experiments. Monte Carlo
simulation has been performed, to follow the propagation of
magnetically converted axions, that is, the created X-rays, near
the solar surface. Secondly, we argue how puzzling solar behaviour
can be the manifestation of axions, revising the so far widely
accepted picture \cite{carlson}, which predicts a bright X-ray
spot from coherently converted pseudoscalars that can show-up only
at the solar disk centre. The reasoning of this previous work is
actually not wrong. We only arrive to different conclusions after
comparing simulation results with solar X-ray observations which,
since they originate from the whole magnetic surface, point
tentatively rather at the photosphere or the (lower) chromosphere
as the axion-to-photon conversion layer than the outer atmosphere
as it was concluded in \cite{carlson}.

\begin{figure*}[htb!]
      \centerline{\includegraphics[height=0.49\textwidth]
                                  {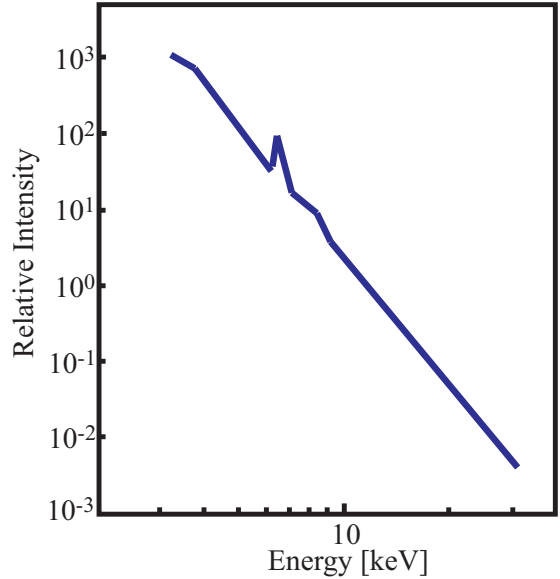}}
  \caption{Qualitative drawing of the usually observed photon analog
  spectrum from solar activities like microflares (see also \cite{hannah08}). A similar
  trend towards lower energies has been observed above 0.8\,keV very recently for the extreme quiet Sun \cite{Syl09}.}
   \label{xrayApprox}
\end{figure*}
Summarizing the previous sections we notice that axion helioscopes
\`{a} la Sikivie \cite{Sikivie:1983ip} (e.g. the running Sumico
and CAST \cite{Mori98,Ino02,Ino08,Zio05,And07,Arik09}) utilize
strong transverse macroscopic magnetic fields in order to force
radiatively decaying particles, like the celebrated axions, to
coherently transform to photons. This working principle raises the
question as to whether and how it could have escaped one's
attention that axion conversion does happen already at the Sun's
{\it ubiquitous} surface magnetic fields. In fact, there are still
conventionally unexplained solar phenomena, from the largest
flares to the weakest transient or almost steady brightening (e.g.
the hot solar corona), which are surprisingly associated with
X-ray activity and solar magnetic fields. The primary process
behind the X-ray flares, for a cool star like our Sun, is still
missing \cite{benz}. Therefore, we follow such observations
further, as possible candidates of solar axion signatures, which
can be as direct as those earth bound helioscopes search for, but
eventually more sensitive, for whatever reason. In the past, the
various solar X-ray activities may not have been considered of
axion origin, ``simply'' because:
\begin{enumerate}
\item the shape of any measured or reconstructed solar analog
photon spectrum decreases rapidly with energy instead of showing
at least a kind of a bump around 4-5\,keV like in Figure
\ref{axion-flux} (see also Figures \ref{xrayApprox} and
\ref{photonIntensity}), \item the topology of the emitted X-rays
does not resemble a spot-like structure that should be located at
the solar disk centre covering $\sim 2\%$ of the solar disk
\cite{carlson}. If the coherent inverse Primakoff effect occurs
far above the surface, there is a perfect collinearity between the
outstreaming axion and the emerging photon. The axion source
(=solar core), the intervening surface transverse magnetic field,
and the Earth X-ray observer define a straight line \footnote{The
solar disk centre was clearly distinguished in \cite{carlson},
i.e. axion related X-rays from the Sun should show up only near
the disk centre. This is actually contrary to everyday experience
with the solar X-ray data from the active and quiet Sun alike
(Figure \ref{patra}). The reasoning of \cite{carlson} applies to
low rest mass pseudoscalars with a relatively large coupling
constant for a QCD-inspired axion, which are assumed to convert
high in the upper chromosphere or beyond. Such a signature has
been searched for since long time. The scenario of this article
might give a reason for this non-observation. Both approaches do
not contradict each other, since the axion rest mass is a crucial
parameter (m$_{\rm a}$$<10^{-4}$eV/c$^2$ in \cite{carlson}, and
m$_{\rm a}$$>10^{-3}$eV/c$^2$ in this work).}. To put it
differently, such axion related solar X-rays point away from the
Earth, if their conversion place is OFF the solar disc centre,
\item a signature from a dark matter particle candidate should be,
by default, extremely faint.
\end{enumerate}

\noindent Generally speaking, only if the above given three quasi
prejudices against the axion involvement in the X-ray bright Sun
can be overcome, will this revise the picture of our nearest star.
Then, the huge sized solar surface magnetic fields should act,
somehow, temporally as an efficient axion-to-photon catalyst due
to a spatio-temporally occurring parameter fine-tuning. The
question is, however, how does this happen? With Figure
\ref{vac_gas} in mind, a possibility would be if the plasma
density of a sufficiently large volume, which is also sufficiently
magnetized \cite{vanBibber:1988ge}, is occasionally `tuned' to the
axion rest mass, enhancing the axion conversion efficiency. This
could explain, in principle, some of the (transient) solar X-ray
emission, in particular if its intensity shows a $B^2$ dependence,
which seems to be often the case
\cite{taup2005,hoffmann,zioutasv7}.

\subsection{Axion(-like) signatures in solar
observations}\label{signatures}

\subsubsection{The hot corona} \label{hotcorona}
\begin{figure*}[htb!]
      \centerline{\includegraphics[width=0.69\textwidth]
                                  {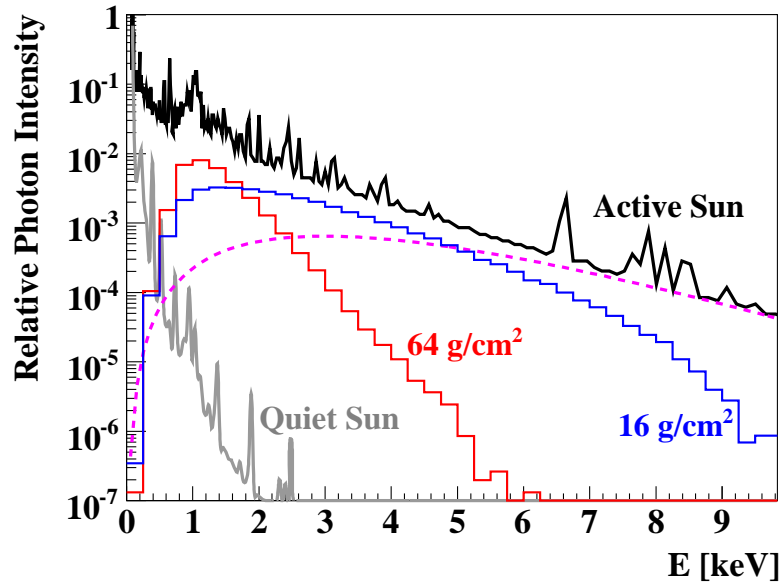}}
  \caption{ Reconstructed solar photon spectrum below 10\,keV from the Active
  (flaring) Sun from accumulated observations (black line). Active Regions are
  associated with strong magnetic fields near the solar surface \cite{Lin08}.
  The dashed line is the converted solar axion spectrum. Two
  degraded spectra due to multiple Compton scattering are also shown for
  column densities above the initial conversion place of 16\,g/cm$^{2}$ and
  64\,g/cm$^{2}$, respectively, which actually agree with the observed spectral
  shape (simulated spectra are not to scale). The same
  interpretation picture could apply to the reconstructed spectrum of the non-flaring
  Quiet Sun at solar minimum (grey line), provided the
conversion occurs deeper into the photosphere. This is also supported by the recent
findings \cite{Lin08} that in the Quiet Sun regions stronger magnetic fields occur
in deeper layers than in the Active Regions. The unknown energy
source of the quiet Sun soft X-ray spectrum reflects the solar corona
problem. Note, the Geant4 code photon threshold is at 1\,keV
(reconstructed solar photon spectra from \cite{per00}).}
   \label{photonIntensity}
\end{figure*}
\begin{figure*}[htb!]
      \centerline{\includegraphics[width=0.79\textwidth]
                                  {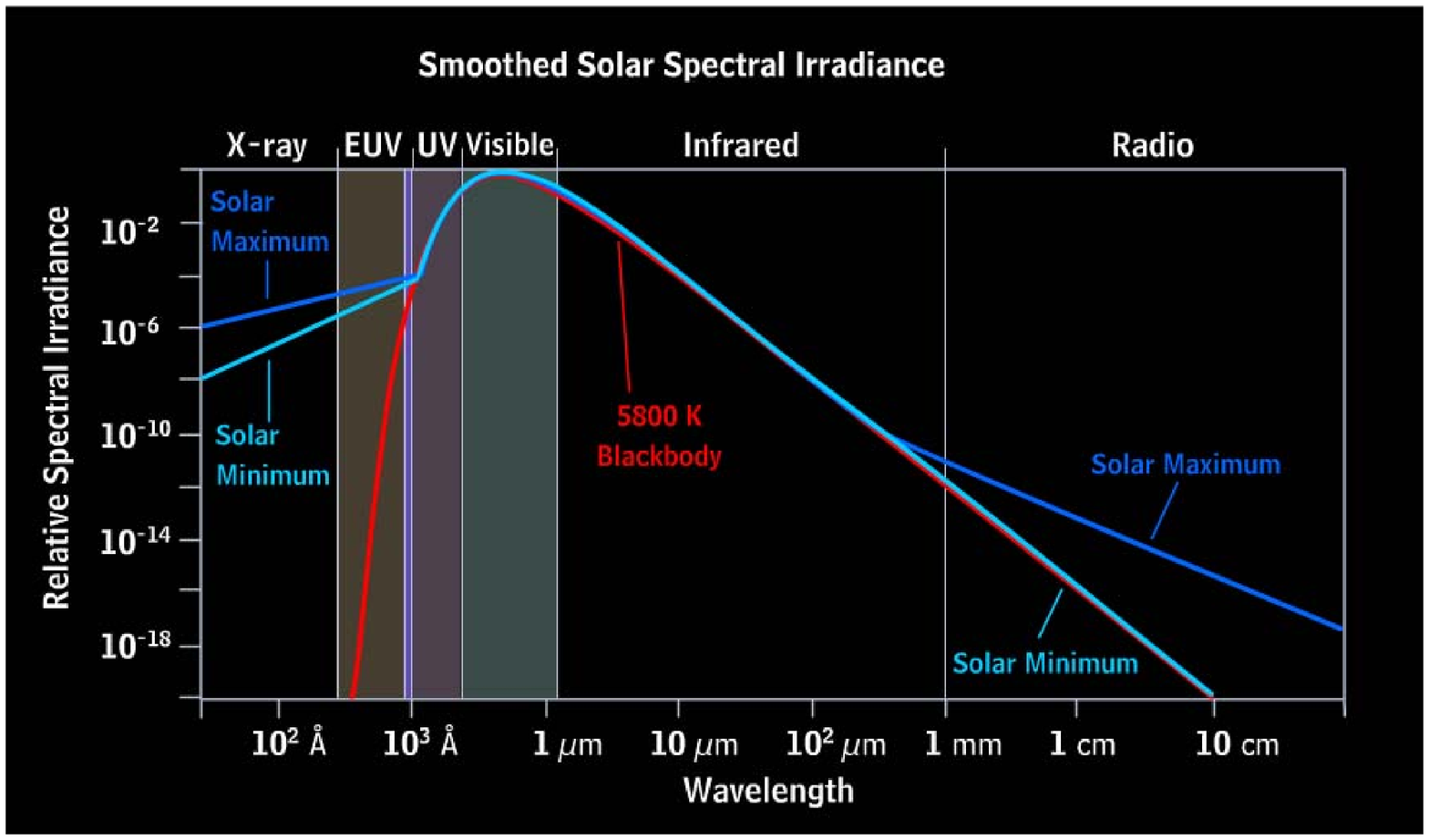}}
  \caption{ The solar irradiance spectrum for two periods in
  the 11-year sunspot cycle. The excess at low wavelengths
  (soft to hard X-rays) above the thermal distribution of
  a solar surface of 5800\,K is conventionally unexpected (see also Figures \ref{photonIntensity} and \ref{transient}).
  Courtesy M.~DiMarco/HAO \& NCAR \cite{irradref}.}
   \label{black}
\end{figure*}
\begin{figure*}[htb!]
      \centerline{\includegraphics[width=0.5\textwidth, angle=-90]
                                  {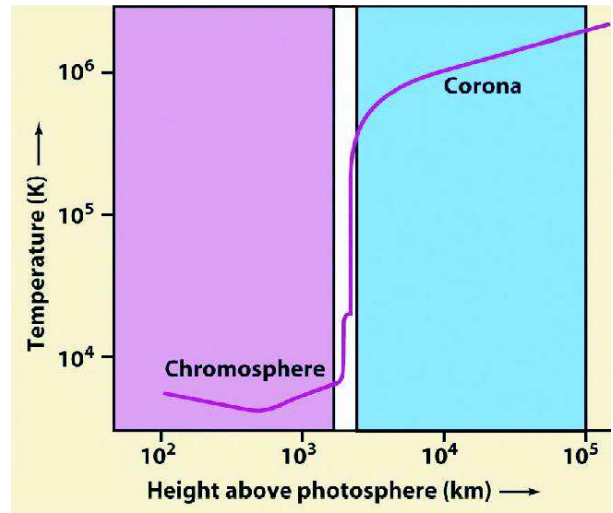}}
  \caption{Atmospheric temperature distribution near the solar
surface. The very existence of the $\sim$\,MK hot solar corona has
challenged astronomers since its discovery in 1939 by Walter
Grotrian \cite{Gro}. Ironically, the Sun's interior is better
understood than its outer atmosphere. Why the temperature is
rising in this way is one of the most challenging questions in
astrophysics. Also the heating of the chromosphere \cite{shibata}
is a long-standing puzzle in solar physics. Courtesy
M.~Weinberg/University of Massachusetts \cite{solcorona}.}
\label{transient}
\end{figure*}

Following the above-mentioned three reasons, even the quiet Sun
X-ray luminosity, which is the famous solar corona problem that
has challenged astronomers since its discovery in 1939 by Walter
Grotrian (Figures \ref{photonIntensity} and \ref{black})
\footnote{ One should note that there is no short of reported
solutions of the solar coronal heating problem. Indicatively, in
the highly valued journal SCIENCE, the almost solution of the old
puzzle has been announced twice \cite{reports}, over a time
interval of a decade, while concluding recently that ``{\it the
question of why the Corona is hot remains unanswered}''.}, should
be naturally excluded from further consideration: the mean photon
energy of the Sun's corona is only about 100\,eV (its temperature
is a few MK), and its spectral shape is a steep exponential one
(Figure \ref{photonIntensity}). In addition, its birth place is
not at all confined near the disk centre, but it covers the whole
Sun (corona). Its intensity is only $10^{-7}$ of the solar
luminosity, but still quite strong and easily observable. How the
Sun increases its temperature within a very short distance from
5800\,K to a few MK (Figure \ref{transient}) is the mystery behind
the solar corona heating.

Previous axion work was motivated by the very steep transition
region (Figure \ref{transient}) separating the chromosphere and
the corona \cite{dilella,zioutas1,zioutas2}. It addressed the {\it
steady} solar X-ray emission as coming from gravitationally
trapped massive axions of the Kaluza-Klein type; their spontaneous
decay near the Sun results to a self-irradiation of the whole
solar atmosphere, which implies an inwardly directed radiation
pressure. Along with X-rays expected to be emitted radially
outwards from converted axions in magnetized places near the
surface (see next), the balance between the two axion-related
radiation pressures could explain otherwise nagging problems with
the Transition Region (see also \cite{judge}), also reconciling
results which contradict robust helioseismological data. For
example, following the axion scenario, the anomalous elemental
abundances can be only a surface effect, evading contradiction
with the inner Sun properties.

Interestingly, the quiet Sun corona is
hotter during solar maximum (T$\sim$2.2\,MK) than during solar
minimum (T$\sim$1.3\,MK) \cite{minmax}. This meets actually the
reasoning of this work, since the magnetic Sun follows the
11-year cycle and magnetically converted axions can in addition heat-up
the atmosphere. More specifically, the solar corona above {\it
non-flaring} Active Regions (ARs) reaches even flaring temperatures
($\sim$4-10\,MK) \cite{Kli95,Nin00,Ko09,Rea09}. Remarkably, the AR
corona can be largely heated to temperatures $>$5\,MK only where the
photospheric magnetic fields are the strongest \cite{Ko09}. In addition, the recent
observation of a hot core (T$>$10\,MK) above a quiescent AR fits this work too \cite{Sch09}.
\subsubsection{Quiet Sun observations with RHESSI}\label{rhessi} It is argued that the reported quiet
Sun X-ray emission \cite{hannahfreiburg} in the 3-6\,keV range,
which is no longer considered as an upper limit (\cite{rhessi2}),
has been explained conventionally as the high energy tail of an
X-ray emitting hot solar plasma. Alternatively, following the
axion scenario of this work, the suggested spectral degradation
for the flaring Sun (see MC simulations below) could be at work
even stronger in the quiet Sun, which implies lower energy
escaping photons due to enhanced Compton scattering (Figure
\ref{photonIntensity}). Recently, it has been observed that the
stronger the surface magnetic field, the smaller the magnetic
effects in the deeper layers and {\it vice versa} \cite{Lin08}.
Then the extrapolation to higher energies will be different
between the assumed thermal distribution \cite{hannah08} and the
one following the squeezed axion-related X-ray spectrum towards
low energies (this work).
\subsubsection{Extreme Quiet Sun observations with SphinX}\label{sphinx} The recently launched space mission
with the SphinX detectors has already provided the first
(preliminary) light curves in soft X-rays (E$_{\gamma}>$0.8\,keV)
for extreme quiet Sun conditions \cite{Syl09}. The observed
power-law spectral shape resembles, at least qualitatively, the
corresponding one given in Figure \ref{photonIntensity} (grey
line). Also this observation fits the axion scenario, and we
consider it -at least- as an additional, independent piece of
experimental evidence from the quiet Sun at its present 11-year
minimum phase. The estimated X-ray luminosity above 0.8\,keV is
$\sim 5\times10^{21}$\,erg/s. Extrapolating the measured spectrum
\cite{Syl09} towards lower energies (grey line of Figure
\ref{photonIntensity}), the total soft X-ray luminosity becomes
$\sim 5\times10^{24}$\,erg/s $\simeq 10^{-9}L_{\odot}$. Even if
these excess X-rays (from both RHESSI and SphinX
\cite{Syl09,hannahfreiburg}) are due to a thermal distribution,
e.g, from a $\sim$10\,MK plasma, the question about the origin of
its actual heating source remains, since this is just the solar
corona problem. Combining the energy distribution and the topology
of the excess events will allow to distinguish between diffuse
\cite{dilella, zioutas1,zioutas2} and transient brightenings (this
work).

\subsubsection{Solar 2D spectra with YOHKOH/XRT} \label{yohkoh}
\begin{figure*}[htb!]
      \centerline{\includegraphics[width=0.79\textwidth]
                                  {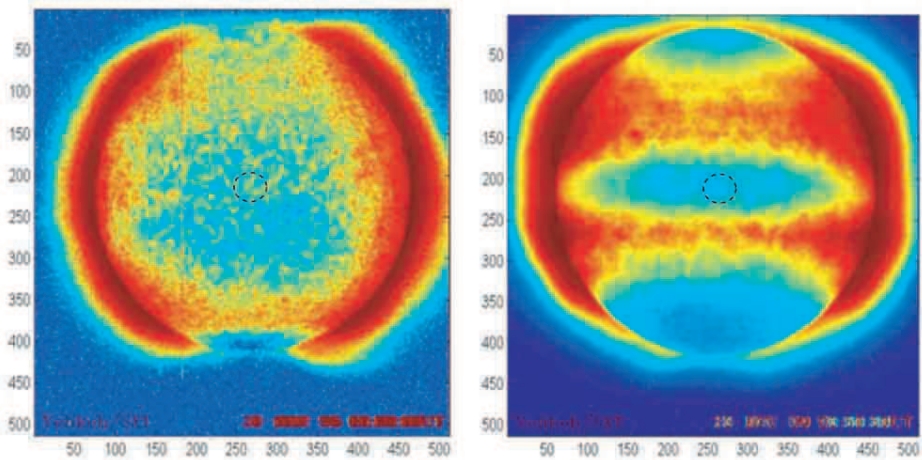}}
  \caption{Solar images at photon energies from 250\,eV up to a few keV from the Japanese
  X-ray telescope Yohkoh (1991--2001). The following is shown: On the left, a composite of 49 of the quietest solar periods
  during the solar minimum in 1996. On the right, solar X-ray activity during the last maximum
  of the 11-year solar cycle. The drawn circles indicate the region of the expected X-ray brightening spot near the disk
  centre, according to \cite{carlson}. Both images show no sign of an X-ray bright spot at the disk centre (see text and
  \cite{carlson}). On the contrary, most of the X-ray solar
  activity (right) occurs at a wide bandwidth of $\pm$35$^{\circ}$ in
  latitude, being homogeneous in longitude. Note that $\sim$95\% of the
  solar magnetic activity covers this bandwidth \cite{How74} (see also a similar topology for microflares measured with RHESSI
\cite{christe2}). This X-ray topology matches this work for a $\sim$10\,meV/c$^{2}$
solar axion or axion-like particle involvement.} \label{patra}
\end{figure*}

In addition to the mentioned quiet Sun steady
X-ray luminosity \cite{Syl09,hannahfreiburg}, a further enhanced
solar X-ray activity (at all levels) is occasionally observed,
which is spatio-temporally correlated with the solar magnetic
activity, and covers a relatively wide band in solar latitude
(${\pm}35^{\circ}$). Both magnetic and X-ray activity appear
equally in all longitudes between the west and east solar limb
(Figure \ref{patra}), with no trace of an enhanced activity near
the disk centre \cite{carlson}. A solar axion scenario must
account for the surface topology of the X-ray distribution and the
power-law spectral shape; the bulk of the emitted intensity is in
soft X-ray emission, in agreement with this work (see below).
Recent observations by the HINODE mission found that the quiet Sun
consists of a network of horizontal magnetic fields
\cite{centeno}. This makes then also the quiet Sun a potential
axion-to-photon coherent converter, which gives rise to the
observed soft X-rays surface distribution (Figures \ref{patra} and
\ref{photonIntensity}). Taking into account that the observed soft
X-ray luminosity of the extreme quiet Sun \cite{Syl09} is $\sim
10^{-9}L_{\odot}$, there is no need to invent any enhancement
factors as required for the largest flares (see \sref{estimates}).
One can always distinguish, however, between a diffuse
contribution and individual X-ray sources from magnetic places.

\subsection{First estimates} \label{estimates}
If one takes the experimentally derived upper limits for
the axion-to-photon coupling strength (see e.g. \cite{Arik09}), a
solar X-ray luminosity from converted QCD-inspired axions should
be around 20 orders of magnitude below that of the visible Sun
(3.8$\times 10^{33}$\,erg/s),
 i.e. extremely small. Therefore, to explain (large) X-ray
 flares, whose trigger remains a mystery, an enhancement by as much as {$\sim 10^9$} is required,
 even under favourable conditions (see footnote $\|$). In other words, an unforeseen mode of interaction remained unnoticed
 so far, avoiding in this way any contradiction with observationally derived
 couplings, by ADMX, CAST and Sumico. As an example, we mention the concept suggested by
 Guendelman \cite{eduardo} about the axion interaction with magnetic
 field and field {\it gradients}\footnote{At the solar surface, we have the appearance of
 magnetic flux tubes, whose diameter is taken 100\,km and the length some
 thousands of km. The interior magnetic field is taken to be 0.2\,T. These tubes resemble
 the geometry of a solenoid. For this simple geometry recently [after submission of this article]
 E.~Guendelman \cite{privateG} has performed a calculation for the conversion probability of
 axions to photons, if they enter perpendicularly to the axis of the solenoid. Surprisingly,
 the conversion probability for such a solar flux tube is of the order of 50\%. More interestingly,
 if instead of 100\,km, one takes a diameter equal to 10\,km and 1\,km, the conversion probability
 drops to $\sim 10^{-2}$ and $\sim 10^{-6}$, respectively. Needless to say, also these last two
 conversion probabilities are still very large, compared to the estimated yield of $10^{-12}$ (see footnote $\|$).
 Note that there are many such flux tubes at the Sun, which may explain a very wide band of solar X-ray emission.
 In addition, if plasma resonance effects are required, in order to enhance the
 conversion probability, coherence lengths of $\sim$ 1 to 10\,km are actually reasonable to consider
 for the active Sun, taking as reference the static density
 distribution below the solar surface. These estimates may show how the next generation
 axion magnetic helioscopes (and haloscopes?) will look like.}. This field configuration has never been taken into
 account in an axion experiment previously, at least not a priori. Surprisingly,
 solar X-ray activity is associated with places of strong magnetic field
 gradients \cite{zioutasdesy}. Then, both the actual interaction and the very nature of the
 solar exotica in question may be different
 from the -otherwise inspiring- QCD axions. To put it differently, it is not
improbable that we are not yet aware of every process occurring in
the Sun, which goes beyond our standard solar axion picture and we
may have not predicted yet all relevant particle candidates.
Remarkably, as we subsequently show, the actual solar X-ray
spectral shape and emission topology, resemble the standard solar
axion scenario (though strongly modified, only if a near to the
photosphere magnetized layer is the axion-to-photon converter).
All this might explain why such solar axion signatures remained
overlooked for such a long time, as well as why space weather is
unpredictable.

In this part of the article, we focus on the magnetic/X-ray active
Sun. In spite of the otherwise completely unfavourable
observational picture for axions, following the reasoning of
\cite{carlson}, we aim to explain how the solar axion scenario
does apply. This will also allow to explain how such unnoticed
solar signals for axions could leave the X-ray Sun covered with a
veil of mystery since decades. The reader should know that
although we are not solar experts, it is encouraging that solar
X-ray missions like RHESSI and HINODE have implemented axions in
their work \cite{rhessi2}. Then, the insisting enigmatic behaviour
of the Sun leaves room for novel and exotic phenomena. Solar X-ray
observatories including YOHKOH a posteriori are the novel axion
helioscopes in space, thus opening new windows of opportunity.
\section{Isotropic X-ray emission from converted solar
axions (simulation)}\label{xray}
\begin{figure*}[htb!]
      \centerline{\includegraphics[width=0.69\textwidth]
                                  {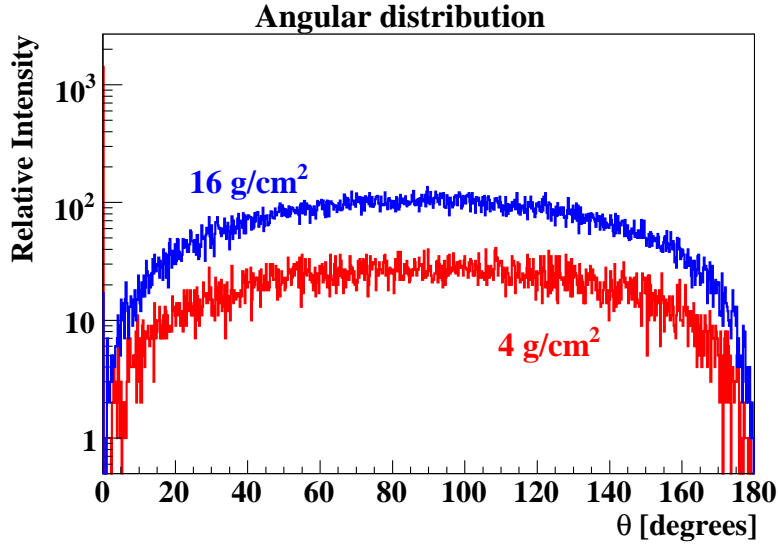}}
  \caption{Simulation with the CERN / Geant4 code. The angular
distribution of X-rays from converted solar axions inside the
magnetized solar surface for two atmospheric column densities
above the axion-conversion place is given. The photoelectric
effect has been inactivated, thus resembling free plasma electrons.
The initial radial axion trajectory direction is taken at
$\Theta$=0$^\circ$. In a vacuum, all solar X-rays from the assumed
inverse Primakoff effect would also escape at $\Theta$=0$^\circ$.
Simulated converted axion events N$_{0}$=16415. Number (N) of not
interacting X-rays at $\Theta$=0$^\circ$: N=1422 or 8.7$\%$ for
4\,g/cm$^2$, and N=18 or 1.1\textperthousand \,\,for 16\,g/cm$^2$.
} \label{theta}
\end{figure*}

A simulated propagation of converted axions to X-rays near the
photosphere has been performed. The novelty is that magnetically
converted axions can be visible even from the whole solar disk for
an Earth X-ray observer. Moreover, the measured analog photon
spectrum can, naturally, be completely different from the original
axion spectrum, being shifted towards lower energies. Remarkably,
all expectations derived from the Monte Carlo simulation fit solar
X-ray observation (Figures \ref{xrayApprox}, \ref{photonIntensity}
and \ref{patra}). In fact, the depth of the actual conversion
region is the only `free' parameter, which can be derived
individually or through combining: the slope (=power law index) of
the X-ray spectrum, the spatial extension of the emitting region,
and the mean photon energy. Statistically, also the obtained
degree of erasure of the initial photon directivity is dependent
on the axion conversion depth in the photosphere (Figure
\ref{theta}).

\begin{figure*}
  \begin{minipage}{0.49\textwidth}
      \centerline{\includegraphics[width=0.69\textwidth, angle=90]
                                  {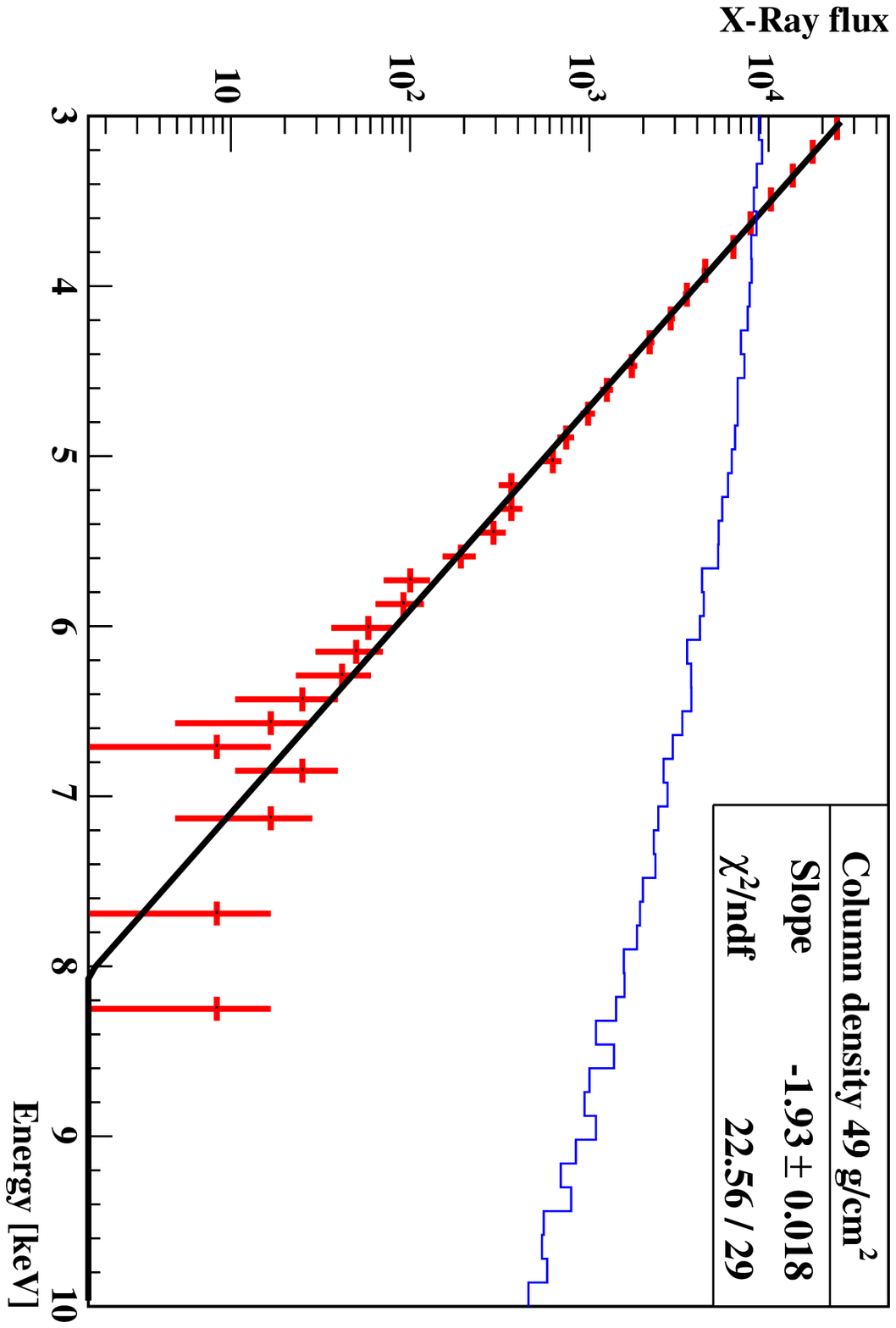}}
  \end{minipage}
  \begin{minipage}{0.49\textwidth}
      \centerline{\includegraphics[width=0.69\textwidth, angle=90]
                                  {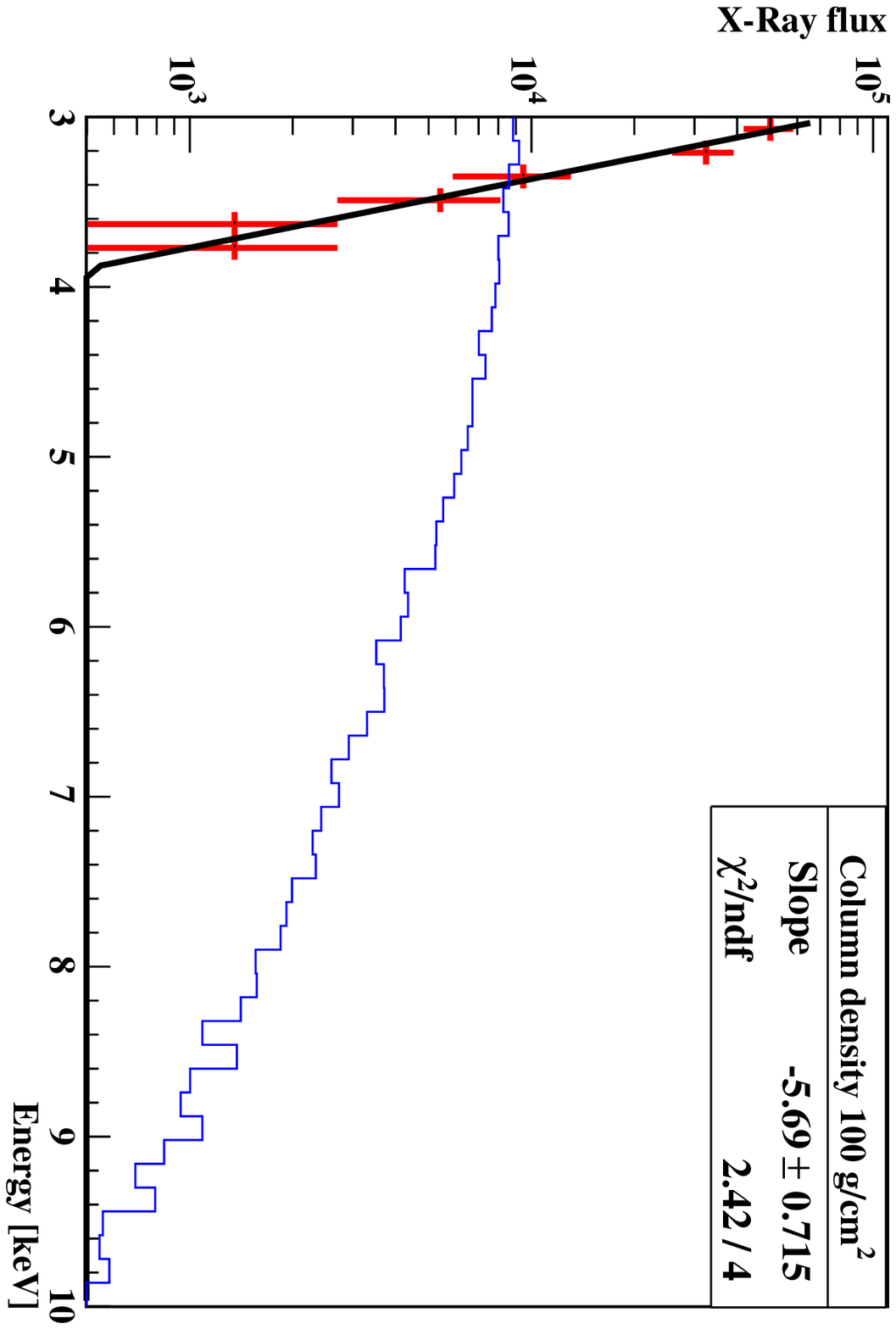}}
  \end{minipage}
  \caption{Simulation of the non-linear energy degradation due to
  multiple Compton scattering of hard X-rays, which start inside
  the photosphere with a solar plasma column density above
  49\,g/cm$^2$ (left) and 100\,g/cm$^2$ (right). The initial
  energy distribution (thin line histogram in blue) is that of the solar
  axion spectrum (Figure \ref{axion-flux}). The strong change
  of the initial analog spectrum depends on the photon's random
  path. The steepness of the distributions depends sensitively
  on the initiation depth, i.e. on the axion-to-photon conversion
  place, where the otherwise unexpected X-rays are assumed to be
  emitted radially outwards inside the relatively cool photosphere
  (T$<$10\,000\,K). For the same density, the depth can
  vary in the dynamic Sun. Surprisingly, these otherwise colourless
  spectral shapes reflect solar observations (Figures \ref{xrayApprox}, \ref{photonIntensity}), with
  the emitted photon spectra dominating exponentially towards lower
  energies (see also Figure \ref{allDensities}).}
 \label{columndensities}
\end{figure*}
\begin{figure*}[htb!]
      \centerline{\includegraphics[width=0.69\textwidth]
                                  {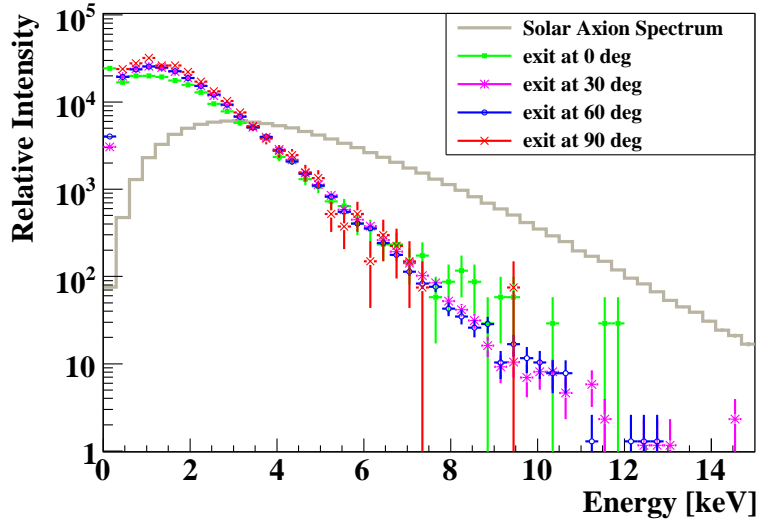}}
  \caption{The same as in Figure \ref{columndensities}, assuming
  the X-rays are created at a depth of 500\,km into the
  photosphere. This Monte Carlo simulation shows that the shape
  of the expected analog spectrum does not depend on the relative
  angle of observation, which is actually counterintuitive. Note, that
  the Geant4 photon threshold is at 1\,keV, and therefore the
  turndown around $\sim$1\,keV is an artefact. }
\label{allDensities}
\end{figure*}
\subsection{A qualitative estimate}
In order to divert an X-ray in the $\sim $1-10\,keV energy range
from its collinear direction of propagation with the converted
axion trajectory, one or even the only possibility is via the
isotropic Compton scattering. For this to happen, a thick plasma
above the axion conversion place is required, suppressing the
photoelectric effect. In fact, also a self-photoionization can
start with converted axion-like exotica near the magnetized solar
sub-surface, which can irradiate and transform an overlying
neutral layer to a plasma. Appropriate environmental conditions,
e.g. plasma density resonance effects over extended regions, can
considerably enhance the axion interaction. \footnote{If converted
axions are the only source of ionization, this might last for some
time. The estimated time to photoionize, e.g. $\sim $2\,g/cm$^2$
above the flare trigger place is of the order of 10$^3$\,s,
assuming an integrated axion originated solar X-ray surface
brightness of 10$^{-2}L_{\odot}$ to be the energy source. This
requirement is in principle possible, even though extreme
(however, the appearance of a large solar flare is also an extreme
and relatively rare event). It cannot be excluded that some other
conventional reaction mechanism is in synergy, like the celebrated
reconnection of opposite magnetic fields. This implies a magnetic
field gradient across the neutral line, which is, surprisingly,
linked to solar X-ray activity. In the quiet Sun, only the tiny
layer below the transition region to the deep photosphere is
actually not fully ionized. Moreover, even for large X-ray flares,
the surface brightness does not actually exceed the quiet Sun
luminosity. The flare region is heated up to 10-30\,MK
\cite{zioutasdesy,rhessinugget}, remarkably close to that of the
core $\sim$ 700\,000\,km underneath. Within the so defined solar
axion ID, the X-ray sun reveals its otherwise hidden face.}

In any case, a more or less bright X-ray flaring region, whatever
the trigger process, remains as a fully ionized plasma until it
starts cooling down to ambient pre-flaring temperatures. X-rays
from converted exotica undergo Compton scattering with the surrounding
plasma electrons. A scattering probability \cite{pdg}, say, of
about 50$\%$, requires an (ionized) column density of
1-2\,g/cm$^2$. Surprisingly, such column densities do exist near the solar surface. For example, some column densities for the
static Sun are \cite{christensen}: a) $\sim$4.4\,g/cm$^2$ at the
surface of the photosphere increasing rapidly underneath, and b)
$\sim$1\,g/cm$^2$ and $\sim$10$^{-3}$\,g/cm$^2$ at +200\,km and
+1000\,km into the chromosphere, respectively. Note that the
plasma density in the solar corona changes dynamically by a factor
of 10-100 at any given time \cite{aschwanden}; this actually holds, at a lower degree, for both chromosphere and
photosphere \cite{hudson}. This implies that the isotropic Compton
scattering of X-rays from converted axions occurring even higher
in the atmospheric plasma can still be quite considerable, i.e. it
might happen even at larger heights than those anticipated for the
static atmosphere.

Consequently, if the actual flaring trigger place is (far) below
the Transition Region (Figure \ref{transient}), the initially
radially and outwardly emitted X-rays from outstreaming and
converted axions can keep the intervening neutral gas above
ionized. It is simply this plasma above the actual flaring trigger
place, which acts as the intervening Compton X-ray scatterer.
Thus, a kind of a dynamic `solar surface effect' can be at work,
whose thickness is only of the order of 1000\,km across the solar
surface. Then, it is this thin layer underneath and above the
photosphere surface, which is `distinguished' within our axion
approach, since it allows for a self-tuning that enhances the
axion conversion, provided m$_{\gamma}$=m$_{\rm a}\sim
10^{-2}$eV/c$^2$. The disk centre region for axions with such a
rest mass is no more peculiar as it was concluded for very light
pseudoscalars in \cite{carlson}. In the past $\sim$13
years, the conclusions from \cite{carlson} might have been
misleading the axion ID in solar X-rays. Axions with
m$_a\gtrsim10^{-2}$eV/c$^2$ result finally to an isotropic X-ray
emission, due to the intervening Compton scatterer, which makes
the whole magnetized solar disk a potential axion-to-photon
converter, visible to an outside observer (Figure \ref{patra}).

\begin{figure*}
  \begin{minipage}{0.49\textwidth}
      \centerline{\includegraphics[width=0.99\textwidth]
                                  {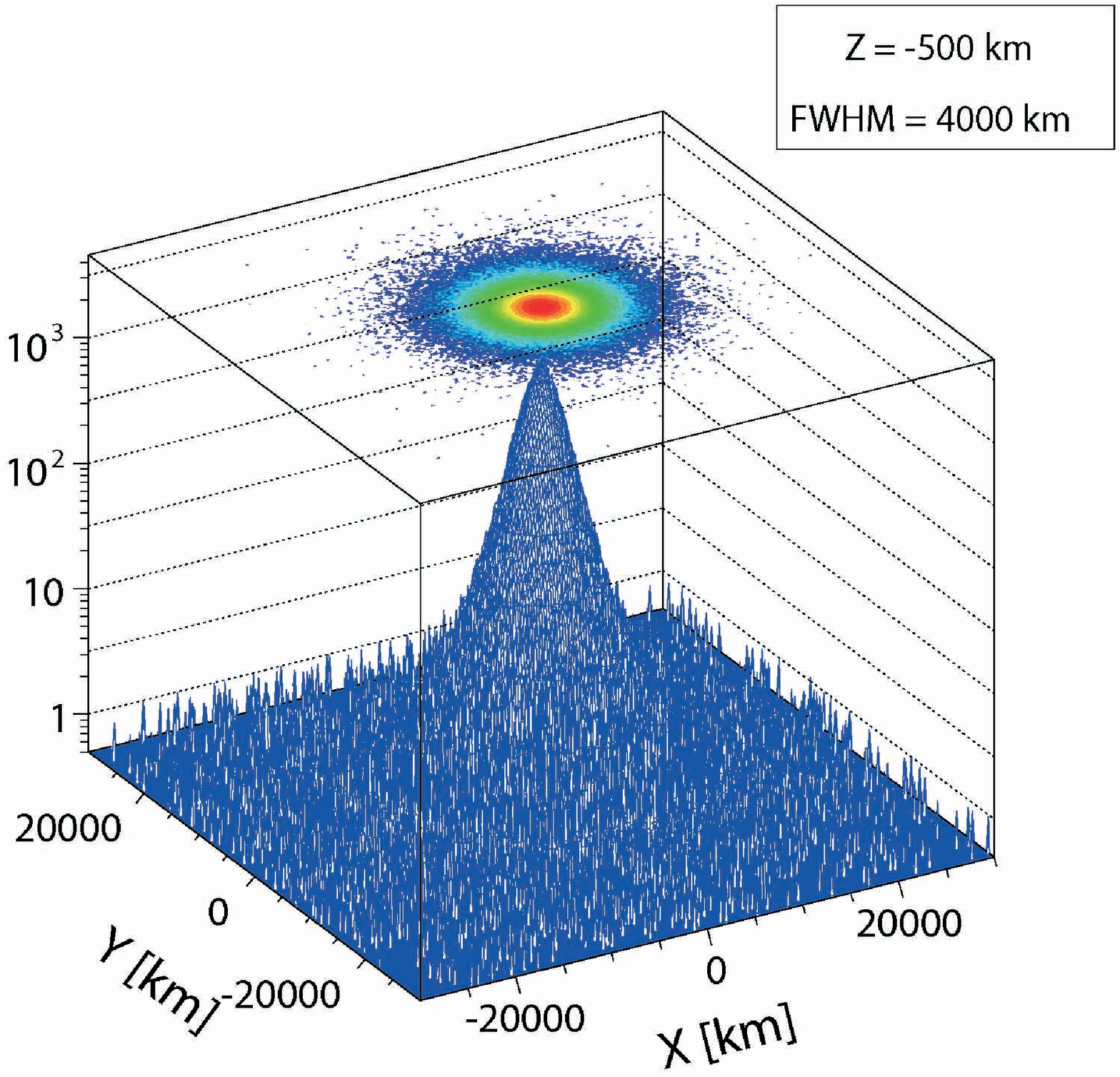}}
  \end{minipage}
  \begin{minipage}{0.49\textwidth}
      \centerline{\includegraphics[width=0.99\textwidth]
                                  {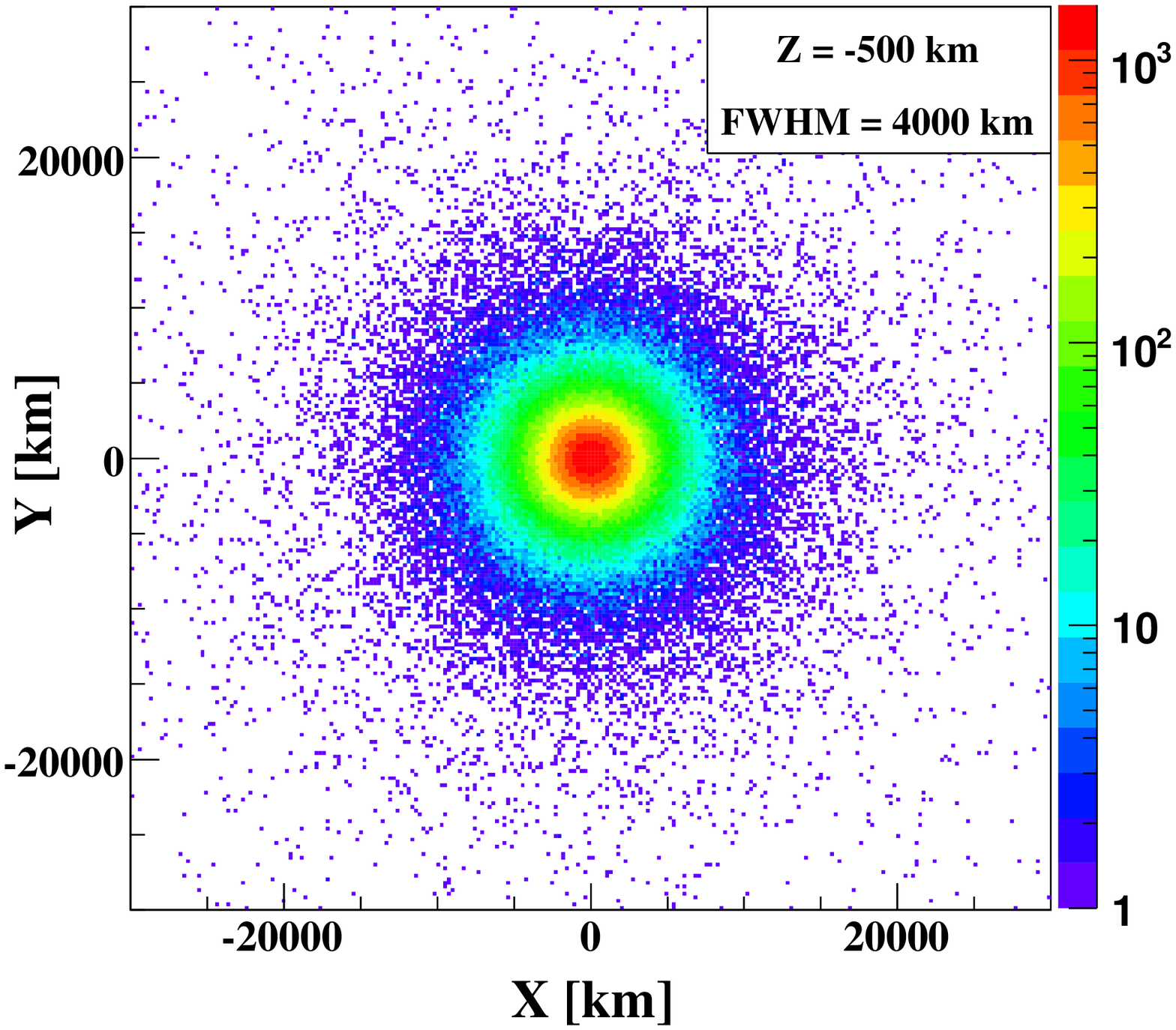}}
  \end{minipage}
  \caption{2D and 3D shower development simulation of an initial
  pencil-like beam of hard X-rays that started at a depth of
  500\,km underneath the solar surface. Their initial energy distribution
  resembles that of the solar axion spectrum (Figure \ref{axion-flux}).
  When they escape from the solar surface after many Compton scattering
  in a random walk, they give rise to a characteristic lateral distribution, which resembles that of a microflare (see \cite{Nin08}).
  Combining this information with their spectral shape (see
  Figures \ref{columndensities} and \ref{allDensities}),
  the initial place of the X-rays can be constrained.}
\label{distributions}
\end{figure*}
\subsection{Monte Carlo Simulation}\label{mc}
In the simulation with the CERN Geant4 code, the photoelectric
effect was inactivated in order to mimic the propagation of
X-rays in a thick plasma ($>$ few\,g/cm$^2$), which after
(multiple) scattering and a `random walk' escape into free space.
The derived isotropic X-ray re-emission from this simulation for
two column densities (Figure \ref{theta}) supports quantitatively
the basic idea behind this work. This means that the information
about the direction of the initial axion trajectory, which is
transferred entirely to the first photon emerging in the
coherent inverse Primakoff effect, is already erased with the very
first Compton scattering.

In addition, in the same Monte Carlo simulation, we have also
studied the photon energy degradation. The next surprising results
are shown in Figures \ref{photonIntensity}, \ref{columndensities}
and \ref{allDensities}. The original X-ray spectrum undergoes a
tiny but non-linear energy redshift after each Compton scattering
due to the energy dependence defined by the Compton kinematics.
This is the reason behind the resulting power law spectral shapes,
as they are given in Figures \ref{columndensities} and
\ref{allDensities}, being strikingly similar to those observed
from active and quiet Sun alike (Figures \ref{xrayApprox} and
\ref{photonIntensity}).

To cross-check the suggested axion scenario, Figure
\ref{distributions} shows the statistical 3D and 2D `shower'
development accumulating many X-ray events; their initial energy
distribution was taken to follow the shape of the standard solar
axion spectrum (Figure \ref{axion-flux}). Interestingly, the
steepness of the spectrum of the escaping photons depends
critically on the depth into the photosphere at which the
axion-photon conversion originates. The plasma density at the
initiating conversion place provides input on the axion rest mass
(Figure 3). In addition, within the axion scenario of this work,
the spatial extension of the escaping X-rays (Figure
\ref{distributions}) allows as well to independently derive the
depth at which the propagation of the initial X-rays have started.
At the current stage of this work, it is also encouraging that the
derived surface size (Figure \ref{distributions}) fits
observations with microflares (Figure 3 in \cite{Nin08}). This
additional feature in favour of the axion scenario supports this
work, even though no firm conclusion can be derived on the role of
the magnetic field $B$, or the magnetic field gradient $\partial
B/\partial z$, or both.

\section{Conclusions and Outlook for the Future}
The present status and the performance of the two operational
state-of-the-art axion helioscopes (Sumico and CAST), including
their possible future upgrades are discussed in this work. Both
have the potential to {\it directly} detect solar axions or other
exotica with similar properties in a mass range below
$\sim$1-2\,eV/c$^{2}$. In the next 2-3 years, they will provide at
least the best limits for the interaction of axions with matter,
if: a) the incoherent Primakoff effect is the main source of axion
creation inside the hot solar core, and b) axions all stream
freely out of the Sun isotropically, reaching the earth bound
magnetic helioscopes when they are pointing at the Sun.

In addition, we studied whether the working principle of the axion
helioscopes applies already near the dynamic photosphere, but was
unnoticed before. Since magnetic fields are ubiquitous in the
outer Sun, and, partly -if not entirely- permeating the inner Sun,
this suggested to reconsider solar data from the solar axion point
of view of the present work. This has been done, arguing in
particle physics manner, with the help of a Monte Carlo simulation
of both the active/flaring and the quiet Sun. The considered solar
observations can be reconciled with the axion scenario taking
place near the solar surface. The derived rest mass for the
axion(-like) particle is m$_a\sim$10\,meV/c$^{2}$, assuming that
plasma resonance effects are at work near the photosphere. Then
the photosphere itself or the near (lower) chromosphere is the
actual catalyst for the axion conversion.

We note that magnetic fields are in the conventional picture somehow
 the energy reservoir for X-ray activity, but, in the
axion scenario they are the self-catalyst for the axion-to-photon
oscillation to occur, with the energy resources being
outstreaming axions from the hot core instead. In reality,
conventional and new physics might coexist, being complementary
and not excluding each other. This makes it certainly more
difficult to disentangle the axion contribution as an explanation
of the solar activity \cite{zioutasdesy,zioutascern}.
\begin{figure*}[htb!]
  \begin{minipage}{0.49\textwidth}
      \centerline{\includegraphics[width=0.69\textwidth]
                                  {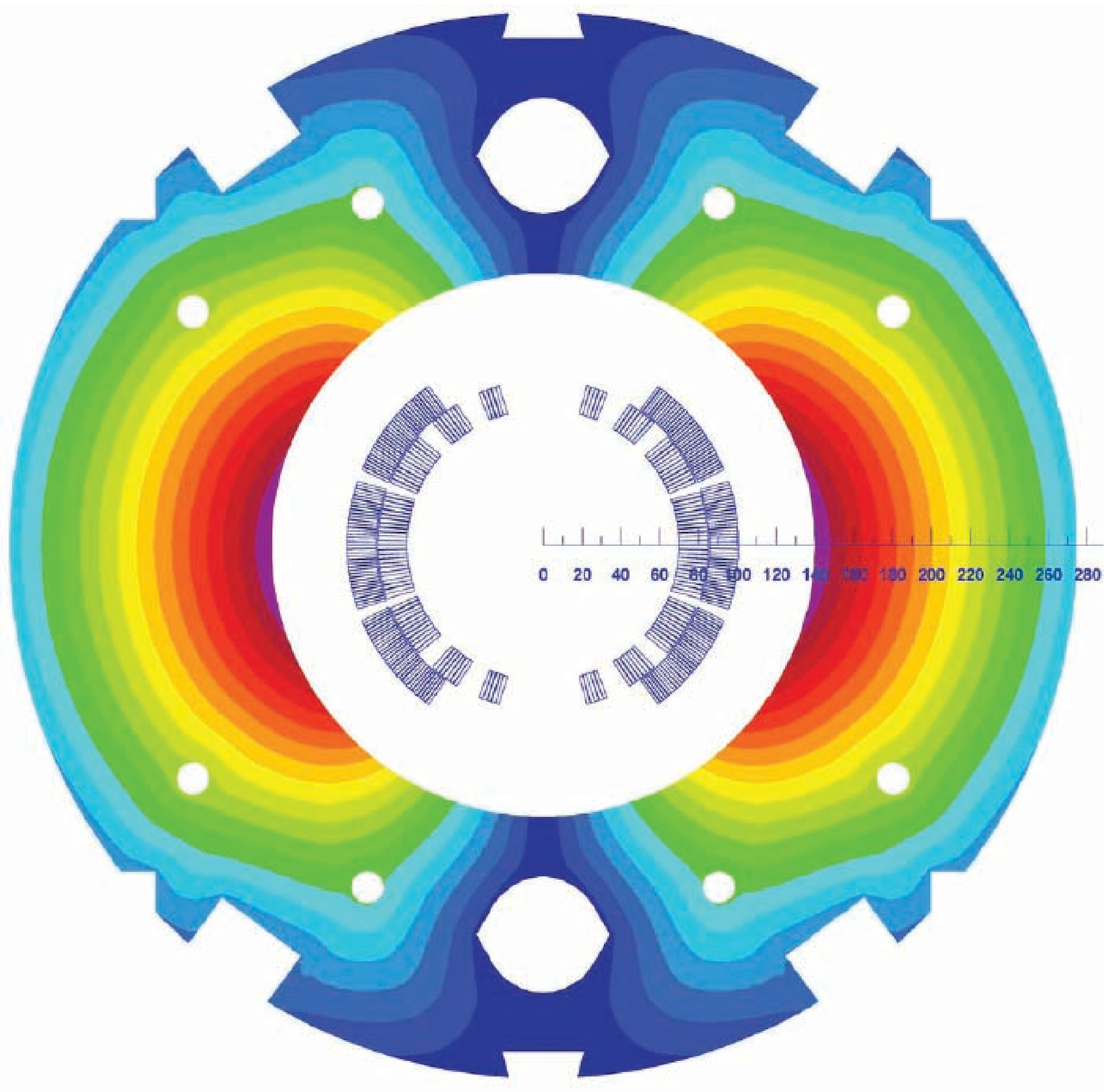}}
  \end{minipage}
  \begin{minipage}{0.49\textwidth}
      \centerline{\includegraphics[width=0.69\textwidth]
                                  {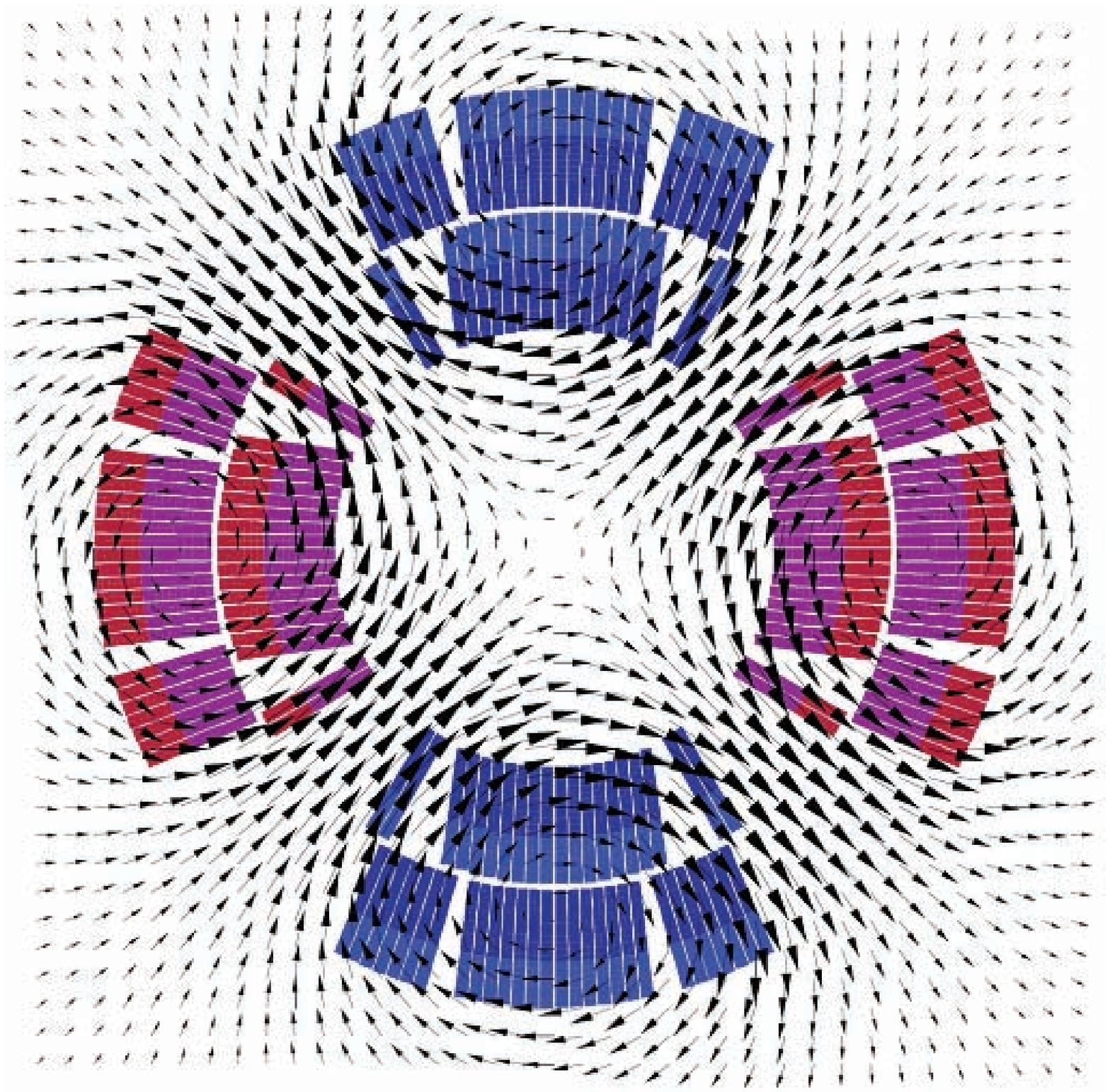}}
  \end{minipage}
  \caption{ New magnets for an upgraded axion helioscope. Left: a
  magnet dipole, 16\,m long, with 9\,T and 140\,mm aperture. Right: a quadruple magnet with a maximum magnetic field of
  $\sim$10\,T and magnetic field gradient up to $\sim$2.5\,T/cm (see text).
  Courtesy: Stephan Russenschuck / CERN.}
 \label{magnets}
\end{figure*}
The magnetically created hard X-rays from axions undergo near the
photosphere multiple Compton scattering, following an outwards
random walk, whose non-linear energy loss dependence changes
drastically the original spectral shape. Its properties like mean
energy, steepness, lateral size, etc. have almost nothing in
common with the conventionally expected solar axion picture, at
least not at first glance. However, the striking similarities of
the simulated features with the directly measured or the
observationally reconstructed ones (Figures \ref{xrayApprox},
\ref{photonIntensity}, \ref{patra}, \ref{theta}), cannot be
ignored, because the estimate of the X-ray emission, within the conventional solar axion
scenario, is rather qualitative. After all, such or other exotica in question might
interact via other "channels", and/or their properties do not
match that much the standard axions. A generic example for this to
happen is certainly the suggested interaction of axions with
magnetic field gradients. This scheme is supported by the
solar observations which find that enhanced X-ray activity
correlates not only with magnetic active regions (sunspots), but
also with places with magnetic field gradients \cite{Cui06,Cui07}.

If an overlooked novel mode of conversion occurs, e.g. in magnetic field gradients, this avoids contradiction between the present
limits on the axion coupling strength and the observed level of
solar X-ray emission from magnetized places, since all magnetic
axion detectors use instead dipole fields. Therefore, a future upgrade of present axion helioscopes
implies a larger and more powerful magnet (Figure \ref{magnets}),
lowest noise detectors, focusing devices, and much lower energy
threshold. Inspired by the solar X-ray activity observations, some
first measurements with quadruple magnets (Figure \ref{magnets})
are in place, leaving room for surprises.

\ack{ We would like to thank the referees for the constructive
comments and criticism over this paper; we do believe that
following their recommendations, this paper has gained in clarity
and (scientific) value. We are thankful to the members of the CAST
collaboration, for the use of CAST related results. Similarly, we
also thank professor Makoto Minowa from the Sumico collaboration.
We greatfully acknowledge the support of Biljana Laki\'c and Magda
Lola. One of us (K.Z.) thanks Hugh Hudson for informative
discussions during his short visit at CERN. We thank Tullio
Basaglia from the CERN library for providing promptly most of the
publications used throughout this work. K.Z. thanks CERN for long
years of hospitality and support of all kind. We give credit to
Eduardo Guendelman for allowing us to use the conversion
probabilities he has calculated prior to publication. The support
we have received from the Greek funding agency GSRT is gratefully
acknowledged. This research was partially supported by the ILIAS
(Integrated Large Infrastructures for Astroparticle Science)
project funded by the EU under contract EU-RII3-CT-2004-506222.
 }
\newpage
\section*{References} 

\end{document}